\documentclass{article}
\usepackage{ijcai23}

\usepackage{times}
\usepackage{soul}
\usepackage{url}
\usepackage[hidelinks]{hyperref}
\usepackage[utf8]{inputenc}
\usepackage[small]{caption}
\usepackage{graphicx}
\usepackage{amsmath}
\usepackage{amsthm}
\usepackage{booktabs}
\usepackage{algorithm}
\usepackage{algorithmic}
\usepackage[switch]{lineno}
\usepackage{quoting,xparse}
\usepackage[table, dvipsnames]{xcolor} 
\usepackage[labelfont=bf,tableposition=top]{caption}

\NewDocumentCommand{\bywhom}{m}{% the Bourbaki trick
  {\nobreak\hfill\penalty50\hskip1em\null\nobreak
   \hfill\mbox{\normalfont(#1)}%
   \parfillskip=0pt \finalhyphendemerits=0 \par}%
}

\NewDocumentEnvironment{pquotation}{m}
  {\begin{quoting}[
     indentfirst=true,
     leftmargin=\parindent,
     rightmargin=\parindent]\itshape}
  {\bywhom{#1}\end{quoting}}

\linenumbers

\title{The pop song generator: designing an online course to teach collaborative, creative AI}

\author{
Matthew Yee-king, Andrea Fiorucci and Mark d'Inverno
\affiliations
Goldsmiths, University of London\\
\emails
\{m.yee-king, a.fiorucci, dinverno\} @gold.ac.uk
}

\begin{document}
\nolinenumbers
\maketitle

\begin{abstract}

This article describes and evaluates a new online AI-creativity course. The course is based around three near-state-of-the-art AI models combined into a pop song generating system. A fine-tuned GPT-2 model writes lyrics, Music-VAE composes musical scores and instrumentation and Diffsinger synthesises a singing voice. We explain the decisions made in designing the course which is based on Piagetian, constructivist ‘learning-by-doing’. We present details of the five-week course design with learning objectives, technical concepts, and creative and technical activities. We explain how we overcame technical challenges to build a complete pop song generator system, consisting of Python scripts, pre-trained models, and Javascript code that runs in a dockerised Linux container via a web-based IDE. A quantitative analysis of student activity provides evidence on engagement and a benchmark for future improvements. A qualitative analysis of a workshop with experts validated the overall course design, it suggested the need for a stronger creative brief and ethical and legal content.

\end{abstract}

\section{Introduction}

Although some of AI's ethical, legal and cultural limitations are now beginning to be recognised, we still find ourselves in a boom time for artificial intelligence (AI). 
As we write, Microsoft has just bought OpenAI for over 10 billion dollars, for example, and the whole educational system is working out how to deal with chatGPT\footnote{``it means the end of homework'' - Elon Musk 2023}.
In addition, there is sustained high-profile coverage in the media of the latest advances in applications of AI to creative or artistic content generation. 
For example, we hear of OpenAI's jukebox creating lost Frank Sinatra songs\footnote{https://www.theguardian.com/music/2020/nov/09/deepfake-pop-music-artificial-intelligence-ai-frank-sinatra}, visual artists suing Stability AI\footnote{https://www.cbsnews.com/news/ai-stable-diffusion-stability-ai-lawsuit-artists-sue-image-generators/} and
AI finishing Beethoven's unfinished symphony\footnote{https://www.scientificamerican.com/podcast/episode/beethovens-unfinished-10th-symphony-brought-to-life-by-artificial-intelligence/}.
%
%footnote{"Beethoven's last symphony finished by AI" https://www.dw.com/en/beethovens-last-symphony-finished-by-ai/a-59412362}
As well as these remarkable achievements, creative-domain AI technology is rapidly becoming more accessible, and this is already transforming a range of practices in the creative industries\footnote{how-ai-is-transforming-the-creative-industries?, The Economist, April, 2021}. 
%
% As experienced practitioners of using AI in music performance ourselves, 
% %
% and long-term educators, teaching AI in higher education, running programmes both on campus and online, we are keen to explore the possibilities of state-of-the-art AI technology. 
% %

We see fantastic opportunities for this technology, from our perspective as experienced musicians who use AI in their creative practice and educators who have taught AI to diverse groups of students on-campus and online for several decades. 
Through our experiences, we see the potential to design and deliver new courses of all kinds such as MOOCs, online, on-campus, and hybrid degrees that use state-of-the-art AI technology to allow students to develop new creative and technical practices. 
%degree-level programmes that use state-of-the-art AI technology to allow students to develop new creative and technical practices. 
% We are in exciting times because the technology 
% is more easy to access and rapidly becoming more powerful 
% and causing a transformation in creative industries which
% people are excited about 

%
% We think an evidence-based approach to AI course design is needed. 
% We think a new kind of AI course is needed. 
%
%But teaching state-of-the-art AI is hard so new approaches to teaching AI are needed.
%
% Given the multi-disciplinary nature of using AI in creative domains, it is necessary to carefully consider course designs and to evaluate and iterate on those designs taking diverse stakeholder perspectives (from research and industry) into consideration. 
% %
% Right now, most people interact with AIs through simple prompts (for example descriptions of images for Stable Diffusion) but once you go beyond basic prompt engineering, the ways in which AI operates rapidly becomes too opaque for effective use. 
% %
% It is important to overcome this brick-wall of complexity encountered by AI users. Left on the outside, we are concerned that AI users might not be able to deeply engage with the systems or to meaningfully inform their future design. 

% %
% Meaningful engagement with AI in creative contexts - including having input into how AI is developed in the future -  requires a deeper 
% % and more hands-on 
% %
% understanding 
% % of how it works, and how it can be used
% . 
%
Given the multi-disciplinary nature of using AI in creative domains, it is necessary to carefully consider course designs and to evaluate and iterate on those designs taking diverse stakeholder perspectives (from research and industry) into consideration. 
We use the phrase ‘AI-creativity’ to refer to AI systems operating in creative domains and practices. We introduce it in order to be as inclusive as possible, encompassing views where AI can be independently creative, where AI is seen as a creative collaborator and or those that see AI as one more tool for creative expression.

Here we report findings from the development, delivery, and evaluation of a new AI-creativity online course, and make the following contributions:

\begin{enumerate}
\item A detailed description of a new online AI-creativity course, and explanation of decisions made throughout its design process;
%alongside clear justifications over why decisions were made to support further research and development

\item A re-usable method for the quantitative analysis of student learning activity within the course and results for 238 students who have taken the course; and

\item An accompanying qualitative approach that enables the analysis of workshops involving pedagogical and professional experts to ensure our learning outcomes are the right ones, the course design is appropriate, and the learning outcomes are being met.

\end{enumerate}

% We believe such courses will become critical for producing graduates with the creative and technical sophistication to work in the creative industries and many others. 
% %
% We also believe that potential students, and future employers, will want stronger evidence about the ways in which students have engaged in any course (especially an online one), and the timeliness and relevance of the material and learning outcomes.

With these contributions, we provide a framework within which future course designers and practitioners of AI-creativity can design, implement and evaluate their materials and the student experience. 
Through this work, future learners will be better prepared with the creative and technical skills to effectively exploit the transformative potentials of AI in our rapidly-changing creative and related industries. 
%^ transformation of the creative industries happening right now. 

% We provide a range of materials that enable other practitioners in our field to build clearly on the foundations we have developed and set up new approaches to designing and evaluating pedagogical materials in AI-creativity.

% \subsection{Research aims}
% [NOTE: these should possibly be rolled into the contributions above. ]
% The research aims of this paper include the following:

% \begin{enumerate}

%     \item To consider the design of an AI-creativity system which is suitable for teaching purposes, which embodies recent developments in AI-creativity technology and which can bridge the gap between low-level AI topics and research-level AI activity
%     \item To investigate how principles of constructivist, inquiry-based learning can be used to immerse students in authentic, real-world problem complexity in the AI-creativity domain    
%     \item To design an AI course that provides students an opportunity to develop their own creative practices in content production through AI-Human interaction
% \end{enumerate}

% \subsection{Guiding Principles}

% \textcolor{red}{ Then I think you need something on constraints. That it doesn't disadvantage student who are less familiar with creative, that it is green in terms of the power used, that it is flexible in terms of new developments in AI and open source products that can be used, that it provides students a clarity on the limitations of AI that can be discovered through creative practice experiences ....  }

\section{Background}

\subsection{Challenges and opportunities for teaching AI in creative contexts}
 Technical subjects like AI are hard to teach, especially when considering a broader student body of non-science students and end-users with varied backgrounds and different mindsets \cite{6595270}. 
Educators commonly employ deductive methods to teach AI,  which one might characterise as ``studying a large body of pre-existing technical knowledge and learning how to apply it deductively to constrained problems designed to test this knowledge" \cite{yee2017steam}. For example, this approach is apparent in the defacto textbook for teaching AI \cite{russell2021artificial}. 

But we are not sure deductive pedagogy is the best approach. We do not think it prepares students for the complexity of using AI in the real world, and indeed, the transformative effect it will have within `Creative Industries 4.0'\cite{lee2022rethinking}. 
% %
% Our view is that this deductive approach does not effectively prepare students for the transition to dealing with the complexity and impact of AI systems and the range of problems
% %
% - both subtle and extreme - 
% % 
%  that arise when deploying AI systems in the real world. Particularly if those problems include the challenge of 
% providing students with the confidence and skills required to 
% interact and collaborate effectively with current AI technologies in creative domains when they need to be mindful of so many issues when doing so. 
% %Matthew - Please check above. 
% % MYK: yeah it needs a good citation describing real world experiences working
% % with AI or needs to go as we do not have space unfortunately. 
So instead of a deductive approach, we design our courses based on an inductive, constructivist approach inspired by \cite{piaget1978psychology,papert1980mindstorms} and more recently, \cite{henriksen2014full}.
These educators advocate an exploratory approach which engages with real-world complexity at an earlier stage of learning.

This inductive, learning-by-doing approach is commonly found in the growing number of `creative computing' and STEAM courses \cite{catterall2017brief}
and is becoming more popular in engineering generally \cite{7044277}. The approach has been found to have positive impacts on student outcomes, such as self-efficacy and 
to encourage more effective learning behaviour \cite{magerko2016earsketch,yee2017steam}. 

Teaching AI through the lens of creative activity is a natural continuation of the STEAM approach. 
There is a growing number of courses and even degree programmes which embed this approach for AI education. 
For example, the MSc Data Science and AI for the Creative Industries at University of the Arts London from Grierson and others\footnote{https://www.arts.ac.uk/subjects/creative-computing},
Computational Media Art courses at Hong Kong University of Science and Technology led by Papatheodorou\footnote{https://cma.hkust-gz.edu.cn/about-cma/},
Computational Creativity courses at Queen Mary, London from Colton and others\footnote{https://www.qmul.ac.uk/undergraduate/},
Goldsmiths' Machine Learning for Musicians and Artists online course by Fiebrink\footnote{https://www.kadenze.com/courses/machine-learning-for-musicians-and-artists/info},
the AI for Media, Art and Design course from Hämäläinen and Guckelsberger at Aalto University, Finland\footnote{https://github.com/PerttuHamalainen/MediaAI}
, and the Creativity and AI specialisation on Coursera from Parsons\footnote{https://www.coursera.org/specializations/creativity-ai}
% , and the Neural Aesthetics course at NYU\footnote{https://ml4a.github.io/classes/itp-F18/}. 

\cite{ackerman2017teaching} list opportunities for teaching AI within creative contexts, such as considering systems holistically and developing and understanding one's own creative processes. 
Authenticity is a related and important aspect of STEAM pedagogy - in a systematic review of authenticity in design-based engineering education (a relative of STEAM), \cite{wang2012conceptualizing} states that "The common theme of all the different authenticity definitions is their relation to real-world experiences". 
\cite{wanzer2020promoting} evaluated their music-oriented programming environment EarSketch and found that its electronic music production environment's authenticity was a factor impacting students' desire to continue learning. 

If we wish for students to engage in a deep, 
sustained and authentic way with the range of AI systems that are garnering the high-profile media attention acknowledged earlier, we need to handle several pedagogical and technical challenges. 
The systems are challenging to describe and understand - for example,  GPT3 has 175 billion parameters and is based on decades of development in natural language processing and deep neural networks. 
The systems are resource-intensive to develop and train - OpenAI's jukebox required hundreds of GPUs and many weeks to train \cite{dhariwal2020jukebox}. 
Even if the student has the hardware and resources to run a given system, the datasets and code are not always openly described or available to investigate. 

% You are making the argument about train again 
%
%
%For example, GPT-3's training hardware cost an estimated \$250 million %\cite{ryabinin2020towards}. 
%%Matthew - One more example 
%The systems are hard to deploy and run - OpenAI's jukebox required hundreds of GPUs %and many weeks to train \cite{dhariwal2020jukebox}. 
% You are making the argument about train again 
%
% To address these challenges, the course design presented here has several features that we will describe later. 
%
In the course that we have designed, we have explicitly set out to address some of these pedagogical and technical challenges and hope that the reader will stay with us later in the paper to find out more. 
 
\subsection{Evaluating AI courses}

In 2019, Fiebrink considered a range of approaches to teaching ML to non-technical creative practitioners, noting at the time that ``Little published research examines how to teach ML effectively to any group'' \cite{fiebrink2019machine}. 

 Our investigations have revealed a few examples relating to pedagogy for AI, including its application to creative contexts, but we found limited evaluation.
\cite{ackerman2017teaching} discuss opportunities and methods for teaching AI in creative contexts, but they do not report an evaluation of their suggested methods. 
\cite{touretzky2019envisioning} consider what every child needs to know about AI, but they do not really consider how to teach it to them, or how to evaluate the results. 
Sanusi notes the potential of inquiry-based learning for teaching ML \cite{sanusi2020pedagogies}, but does not explain how to evaluate it.
In 2020, \cite{marques2020teaching} found several published instructional units for teaching ML to children, and they provide an analysis of the approaches involved but not any evaluation metrics. 
In 2021, \cite{friedmana2021image} evaluated a range of different tools which provide access to generative image models to students on a `synthetic art' course, focusing on the systems' creative potential. 

%the main gap we identified in this brief review was real-world evaluation of course designs. We address this limitation by developing and using an evaluation framework involving quantitative and qualitative analysis, reported in section 4. 

% We aim to address some of these shortfalls in this previous work by employing a clear 

% In our approach, our goals are for students both to learn how AI works, and also how to learn to use AI in creative contexts.
%  %
% Furthermore, we are keen to provide techniques for evidencing how students are learning, and how that learning mirrors both current practices and opportunities afforded by AI. 
%

In summary, educators and researchers are building a body of practice and research around STEAM pedagogy and teaching AI in creative contexts. 
Our work continues in this vein and provides a natural extension of what has been developed to date.
The originality of our work is in the technical ambition, pedagogical specificity, and the development of a mixed-methods evaluation framework.

%%%%%%%%%%%%%%%%%%%%%%%%%%%%%%%%%%%%%%%%%%%
%%%%%%%%%%%%%%%%%%%%%%%%%%%%%%%%%%%%%%%%%%%
%%%%%%%%%%%%%%%%%%%%%%%%%%%%%%%%%%%%%%%%%%%

\section{Course design}
\begin{table*}
\begin{tiny}
    \begin{tabular}{p{2cm}|p{7cm}|p{1.5cm}|p{1cm}|p{2cm}|p{2cm}}
    \hline
    Week title                              & Learning objectives                                                                                                                                                                                                                                                                                                 & Model        & Data           & Technical concepts                                        & Creative activity                                       \\ \hline
%%%%
    Introduction to generative systems
     &
\begin{tiny}   
\begin{enumerate} 
    \item[1.1] Describe the plan and key steps for the AI and creativity case study
\item[1.2] Implement a generative system that can learn a model of a text document and use it to generate more text
\item[1.3] Using examples from the literature, explain what a generative system is                                                                 
\end{enumerate} 
\end{tiny}
    & Markov model & text           & Auto-regression, statistical models                       & Different inputs, different order                       \\ \hline
%%%    
    Generating lyrics with GPT-2            
    & 
\begin{tiny}   
\begin{enumerate}    
\item[2.1] Describe how self-attention allows for a combination of contextual and sequential data in transformer networks
\item[2.2] Instantiate a pre-trained language generating pipeline using GPT-2 and huggingface
\item[2.3] Explain how the process of fine-tuning works and why it is necessary to fine-tune pre-trained neural network models 
\end{enumerate}
\end{tiny}

    & GPT-2        & text           & Self attention, fine-tuning                               & Prompt engineering, fine-tuning with different datasets \\ \hline
%%%    
    Music composition with MusicVAE         
& 
\begin{tiny}   
\begin{enumerate}
\item[3.1] Explain what a variational auto-encoder is and give examples of applications for VAEs
\item[3.2] Describe the concept of a latent space and explain why they are important when exploring the capabilities of pre-trained models
\item[3.3] Load and use a pre-trained model to generate multi-track MIDI files    
\end{enumerate}
\end{tiny}
                 
    & Music-VAE    & MIDI and audio & Autoencoders, latent space                                & Exploring latent space via latent vectors               \\ \hline
%%%    

    Singing voice synthesis with Diffsinger 
    & 
\begin{tiny}   
\begin{enumerate}
\item[4.1] Give examples of historical speech synthesis systems and techniques
\item[4.2] Describe the complexity of current generation speech synthesis models
\item[4.3] Using examples from the literature, discuss different aspects of the concept of creativity 
\end{enumerate}                                                                                  
\end{tiny}

    & Diffsinger   & text and audio & Feature processing and model orchestration                & Realism and different input patterns                    \\ \hline
%%%    

    Putting it all together                 
    & 
\begin{tiny}   
\begin{enumerate}
\item[5.1] Using examples from the literature, discuss different aspects of the concept of creativity
\item[5.2] Differentiate between a Skinnerian view on creativity and a Dewian view
\item[5.3] Put together a complete system which can generate lyrics, music and singing, then mix them into an audio file                                      
\end{enumerate}                                                                                  
\end{tiny}

    & All          & All            & Theory of Ai and creativity, Linking all systems together & Working with pop song fragments                         \\ \hline
%%%    

\end{tabular}
\end{tiny}
\caption{Pop song generator case study course design. }
\label{tab:course_design}
\end{table*}

In this section, we will give an overview of the course design that is summarised in table \ref{tab:course_design}.
Before we do so, 
and for the purposes of significance and reproducibility, we are keen to be explicit about the  list of requirements and constraints for the course, that guided our choices. 
Some of these have been explained in the background section and some will be introduced here for the first time and subsequently explored.
We are guided and constrained by the following: 

\begin{enumerate}
\item Contains AI-creativity content allowing for authentic, engaging creative and technical activities
\item Covers multiple AI models whose outputs need to be produced into an output cohesive and engaging final item 
\item Students should not require special hardware or be required to spend excessive time on complex setup work which is not linked to course learning objectives
\item Implements best-practice learning design for the Coursera platform and the pedagogical practices developed with Coursera for our online computer science degree
\item Organised into five weeks' content with a total study time of about 25 hours
\item Fits with a set of three other AI case studies in the 20-week, 100-hour undergraduate AI course
\item Suitable for final year CS undergraduates, but readily adaptable for future wider audiences 
\item Can be readily refined and extended to include more extensive case studies and a more explicit investigation into issues such as ethics, appropriation and copyright.

\end{enumerate} 

We considered a range of domains for creative activity, such as images, sound, and text. We eventually landed on the concept of a pop song generator. Generating lyrics, music, and singing voice for a short song provides a  combination of different techniques and models with a clear overall concept. 
We also have expertise in the music domain, which helps us with technical and creative development. 

We present the material as five weeks of content delivered in a 20-week online undergraduate course in artificial intelligence, part of our online computer science degree. So the initial audience was undergraduate (UG) CS students. We plan to expand this audience by making the course available for any learner in a standalone MOOC format on Coursera. 
This content is the final part of an AI course comprised of several AI case studies including 
automated scientific discovery, game-playing AIs, and evolving robot morphologies.

Existing best practices on both the Coursera platform and our online CS degree (launched in 2019 and now with 6000 students) have also influenced the course design \cite{hickey2020drivers}. 
This practice suggests optimal learning design features such as alternating between videos within a 5-15 min length range and short formative quizzes reviewing the video content. 
The course should also contain structured activities that have clear outputs connected to learning objectives, and that can ideally be completed within the platform. 
%We should note that our degree was the first undergraduate degree on the platform, so previous best practice relating to MOOCs is not always applicable. 
%
\subsection{Developing the pop song generator prototype}
%One more strand, and perhaps the most challenging one, was the technical development of the pop song generator system. 
We decided upon three major components for the pop song generator system: a language model to generate lyrics, a symbolic music model for the musical score, and a voice synthesis model to sing. 
\subsubsection{GPT-2 lyric generation}
\begin{pquotation}{GPT-2 + eurovision}
I want to go dancing with your eyes closed. I want you to dance with my arms crossed. I want you to dance with my heartbeat.
\end{pquotation}

The first component used huggingface's GPT-2 implementation for language generation \cite{wolf2019huggingface}. 
We selected huggingface as it provides a mature API and a repository of pre-trained models, providing our students with choice.
With this setup, we can generate language in three lines of Python. 
We also provide students with a huggingface GPT-2 model that we fine-tuned in advance 
% (with a clear explanation to our students) 
with a dataset of Eurovision song contest lyrics. 
Since students would need special hardware to fine-tune, we showed them the process of preparing the dataset and fine-tuning the model in a video. 
We made it possible that students could follow our process if the necessary hardware was available to them. As well as fine-tuning and the basic idea of autoregression, GPT-2 allowed us to discuss transformers and how attention is used to blend context into word embeddings. 
%We demonstrated the fine-tuning dataset, which is likely a more common practice than training a complete model from scratch.
%Students could explore creatively here simply by experimenting with the vanilla, and fine-tuned models with different conditioning prompts. Or, more technically, experienced students could fine-tune the model themselves with their dataset, though this required them to go beyond the instructions in the course.  

%
\subsubsection{Music-VAE MIDI generation}
The second component used Google Magenta's Music-VAE for musical arrangement generation \cite{roberts2018musicvae}. We chose this as the researchers provide a range of online examples built in the Javascript language, which ran straight away in the web browser, and a Python implementation. In the end, we could not operationalise the Python implementation, so we provided students with a minimal tensorflow-js version. This version ran on the command line in the Coursera Labs environment using node.js. We provide more details below about the Labs environment. 
We used the Music-VAE system as a vehicle for the concepts of variational auto-encoders and latent spaces. In particular, we demonstrated how students could creatively explore the latent space by permuting latent vectors. 

\subsubsection{Singing voice synthesis with Diffsinger}
The most technically challenging element was the singing system. During the course development, we realised that converting text and musical scores to singing is a very active, though niche area of research. We used a slightly modified version of an open source implementation by Keon Lee \cite{lee2021diffsinger} of the Diffsinger text to singing synthesis system \cite{liu2021diffsinger}. Diffsinger uses several models in combination to generate and process a series of different features, so we used this to teach students about model orchestration and feature processing. Similarly to the other two systems, the singing system allows students to experiment creatively at several levels. They could pass in the different musical note and lyric sequences or dig into the code and adjust the lower-level control parameters for the synthesis system, such as pitch and amplitude modulation. 
\subsection{Running the pop song generator on Coursera}
\begin{figure}
    \centering
    \includegraphics[width=\linewidth]{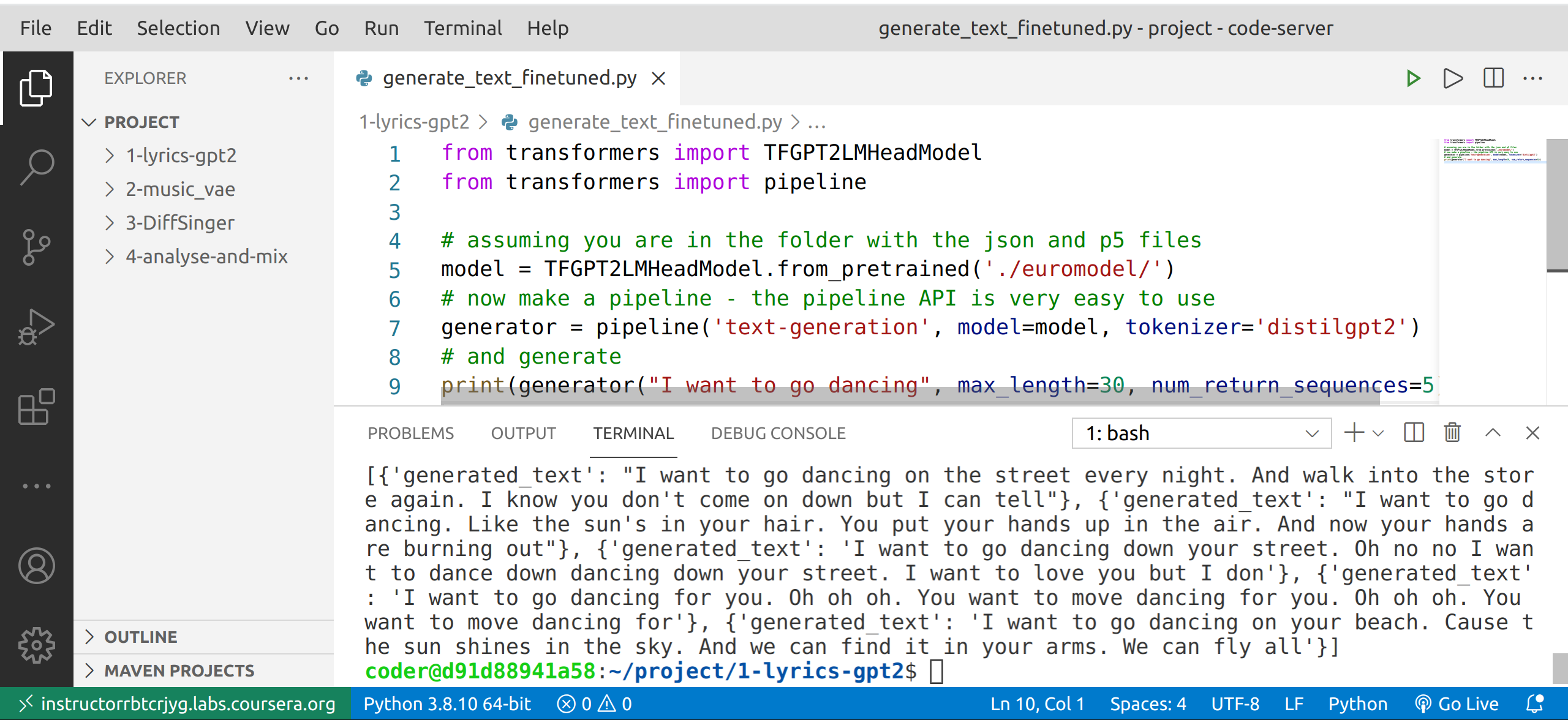}
    \caption{The Coursera Labs IDE running in a web browser. We set this up with all three systems ready to execute. GPT-2 is running in the terminal at the bottom right.}
    \label{fig:ide}
\end{figure}

Coursera provides a docker-based system called Labs where it is possible to run Jupyter notebooks, or any arbitrary web application.
% on the Coursera compute infrastructure. 
The student can access their own instance of the application, effectively their own private server, using their web browser. 
We were keen to enable all three AI systems to run together so that students could move from watching a video to running a state-of-the-art model via their web browser easily. 
Figure \ref{fig:ide} shows a screenshot of our Coursera environment running a GPT-2 model. 

The default setup for Coursera Labs has limited power available with 4GB of RAM and two CPU cores and no GPU acceleration and for security reasons, the environment has minimal internet access. 
Therefore, we ensured all models could run in inference mode (as opposed to training mode) in a reasonable time without accessing blocked parts of the internet. 
The final setup consisted of a virtualised Linux environment providing a browser-based IDE (Visual Studio Code), embedded terminal, and all required software and models installed and ready to run. 

We appreciate that setting up a local Python environment, learning how to install packages, working around clashing versions and incompatibilities 
and so forth is an important part of becoming a machine learning engineer. But as educators, we must decide what we wish to teach in a given context. For this context, our learning objectives were oriented more toward understanding and using the systems than installing them. 

%The most challenging part of the lab build was configuring huggingface to access local copies of the models instead of downloading them from their repository. Once we had solved this, 
We found that the GPT-2 models were fast enough in inference (generative) mode, typically generating output in a few seconds once the models were loaded. 
Music-VAE ran in node.js from the Lab terminal, taking around 30 seconds to generate a MIDI file. 
Diffsinger ran in about a minute, as our customised version needed to render twice to achieve the correct length of audio for the song.  
The full implementation of our pop song generator is openly available in a GitHub repository\footnote{anonymised}.

\subsection{Producing the teaching materials}
%
%
% \begin{figure}
%     \centering
%     \includegraphics[width=\linewidth]{vid2.png}
%     \caption{A scene from the case-study videos.}
%     \label{fig:vid}
% \end{figure}
%
%
Once we had the pop-song software working on Coursera, we produced the actual teaching materials, including wide-ranging content such as videos, multiple choice quizzes (MCQs), workshops in the Coursera Labs environment, peer reviews, and asynchronous discussion activities. 
Videos were produced using a one-person-operated video studio developed during the pandemic. 
The video studio is equipped with hardware and custom software, making it possible to create a live video edit with multiple camera shots and a green screen.
%A screenshot of one of the videos is shown in figure \ref{fig:vid}. 
%We placed a short MCQ after each video to check student understanding. 

Producing the lab worksheets was a significant part of the development work. The worksheets would take students through the processes they had seen in the videos. They would do this inside the virtual lab environment.  
Open-ended challenge exercises encouraged students to go beyond the content in the videos. In particular, we encourage the students to explore the creative aspects of the AI-creativity systems in each course section. For example, they experimented with prompts and the fine-tuned version of the GPT-2 model, permuted latent vectors with the Music-VAE model and explored low and high-level control parameters in Diffsinger. Table \ref{tab:course_design} summarises some of the creative and technical concepts covered each week. Discussion prompts allow for the lightweight sharing of system outputs and small search and report activities. Peer reviews allow for more in-depth checking of technical and creative progress, graded by students against simple rubrics. Ultimately, we assessed students with a written exam wherein students answer short and long-form theoretical and technical questions. 

%We use discussion prompts for two purposes: to encourage students to share their most interesting creative outputs (typically after completing a lab worksheet) and engage them in theoretical and ethical discussion. 

\section{Analysis}
We have described our `minimum viable product' course design, but is the course design successful? 
%We have explained how the course design addresses several technical and pedagogical challenges, dealing with multiple requirements and constraints. 
%Also, the design aims to engage students in active exploration with the AI systems, and ultimately we would like to consider how to make the material accessible to a broader audience, including creative practitioners. 
We have employed several methods to evaluate the course design against our goals. In this paper, we will present two tranches of our ongoing evaluation: a quantitative analysis looking at `time-on-task' and a qualitative evaluation based on a thematic analysis of a workshop we conducted with AI and creative industry experts. 

\subsection{Quantitative analysis of student activity}

The course has now been taken by 238 students from our UG CS programme on Coursera. The Coursera platform generates extensive data exports allowing for detailed analysis of student activity. For this paper, we have focused our quantitative analysis on the duration of student access to the five parts of the course compared to the degree cohort as a whole. 
We chose duration as it is a well-established proxy metric for student engagement - Wong et al. state that "Time-on-task has long been recognized to be a significant variable that is correlated with learner engagement as well as a predictor of learners' achievement" \cite{wong2018modelling}. 

\subsubsection{Method}
We extracted log files containing timestamps for every student's access to the course items, which we refer to as `hits'. We can then organise the hits into chronological sequences for each student on each course and compute the intervals. We can then sum the intervals to find the total time spent on a course. We exclude intervals greater than 4 hours as we assume those indicate students ending their study session. We also measured the time spent in lab activities in the pop song AI content and labs in the entire BSc degree. We organised the times into six groups: BSc CS as a whole (5842 students) and the five weeks and labs of the AI course (238 students). Thus our research questions for this analysis are:

\begin{enumerate}
    \item How long do students spend per week in the pop song generator course compared to the BSc CS as a whole?
    \item How long do students spend in the lab activities in the pop song generator course compared to the BSc CS as a whole? 
\end{enumerate}

\subsubsection{Results}

\begin{figure}
    \centering
    \includegraphics[width=0.5\textwidth]{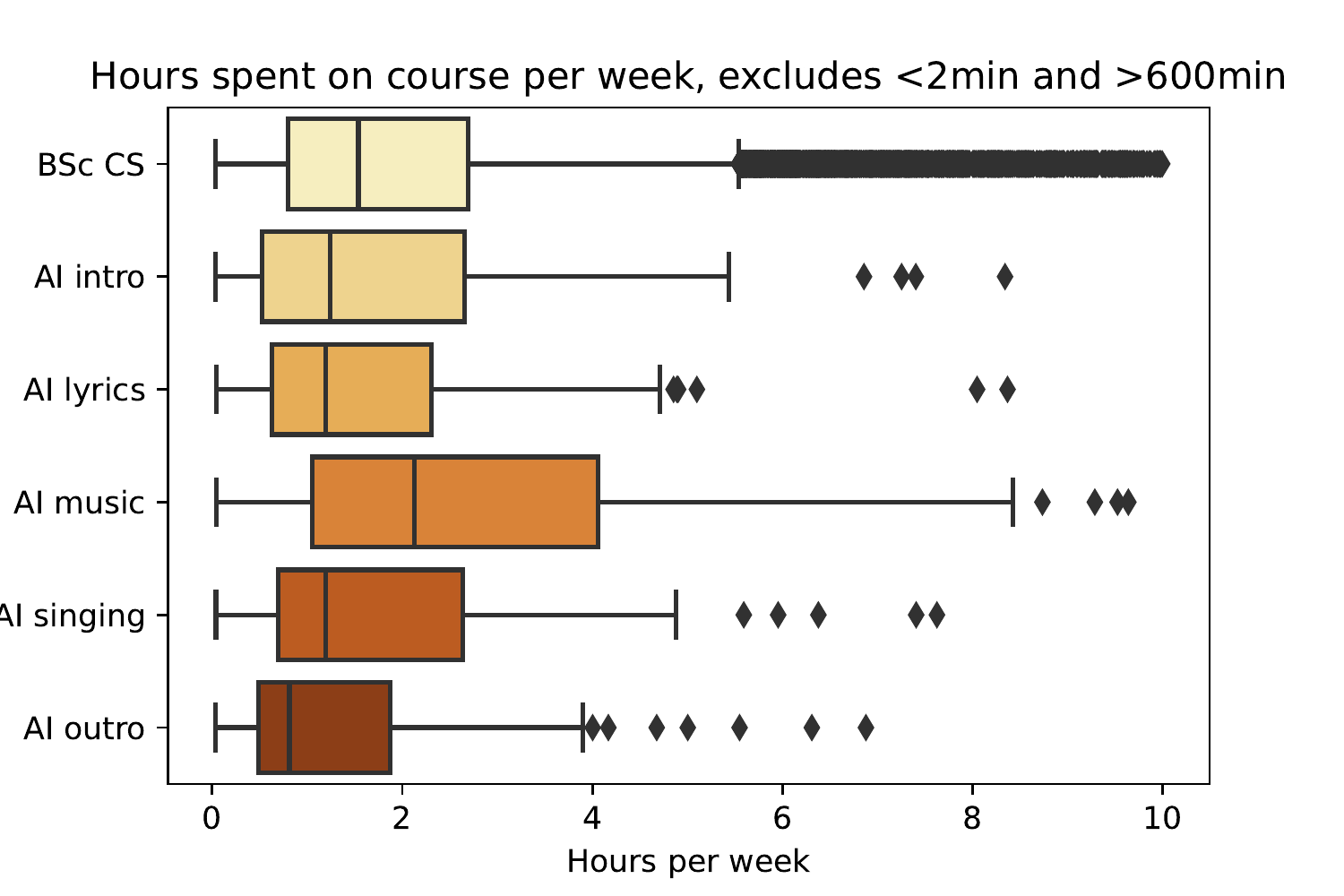}
    \caption{Range of times spent per week on the course and the degree as a whole for comparison. The main bars show the mean and deviation, and the points to the right show outliers. The variation between BSc and AI is significant, except week four is similar to BSc. }
    \label{fig:weekly_time}
\end{figure}

\begin{figure}
    \centering
    \includegraphics[width=0.5\textwidth]{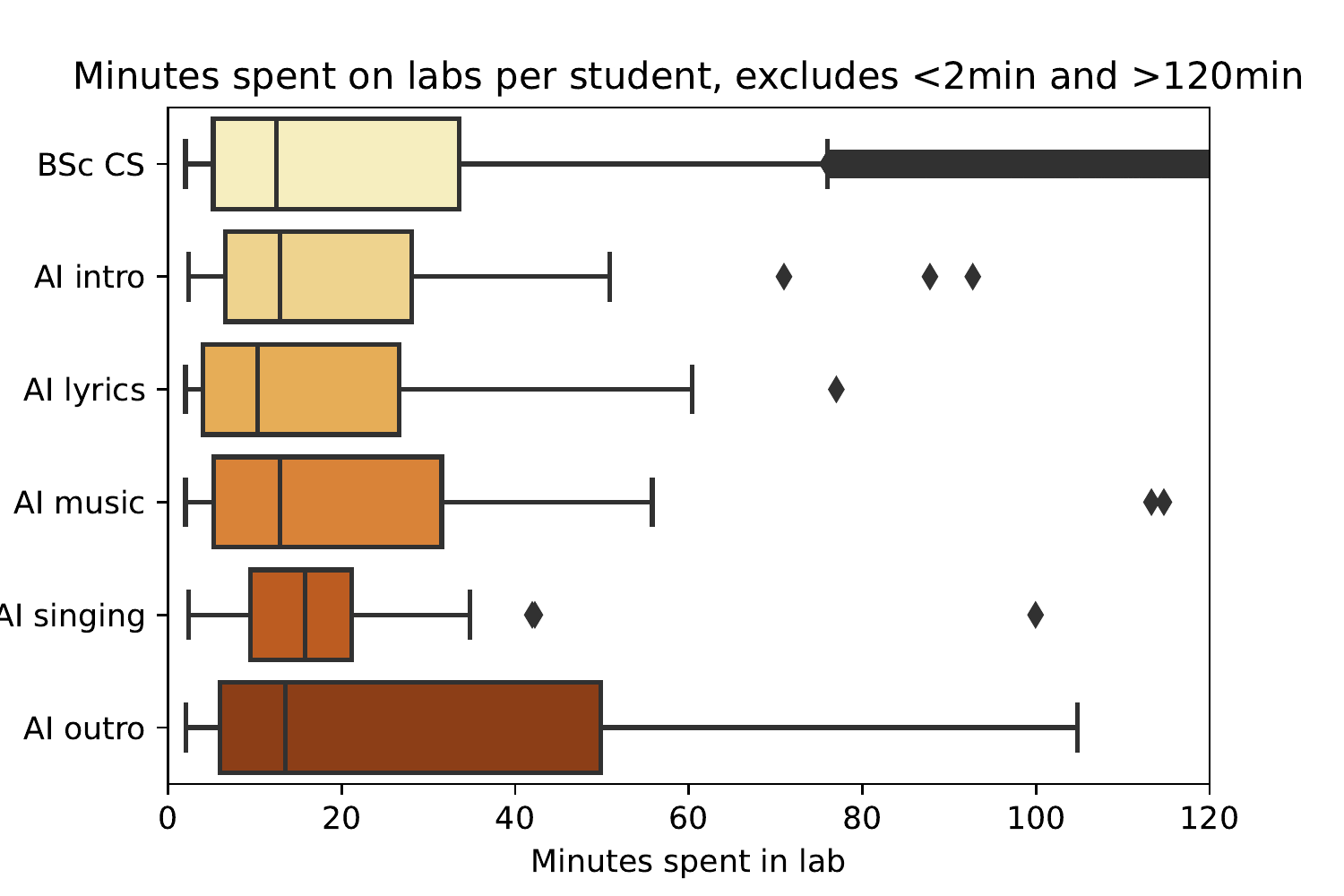}
    \caption{Range of times spent in labs in the degree as a whole and each week of the AI-creativity course. }
    \label{fig:lab_time}
\end{figure}

% \begin{table}[]
% \begin{tabular}{l|l|l|l|l|l|l}
% \hline
%        & BSc CS & AI 1 & AI 2 & AI 3 & AI 4 & AI 5 \\ \hline
% BSc CS & 0      & 0.02 & 0.02 & 0    & 0.06 & 0    \\ \hline
% AI 1   & 0.02   & 0    & 0.86 & 0    & 0.82 & 0.02 \\ \hline
% AI 2   & 0.02   & 0.86 & 0    & 0    & 0.86 & 0.01 \\ \hline
% AI 3   & 0      & 0    & 0    & 0    & 0    & 0    \\ \hline
% AI 4   & 0.06   & 0.82 & 0.86 & 0    & 0    & 0.01 \\ \hline
% AI 5   & 0      & 0.02 & 0.01 & 0    & 0.01 & 0    \\ \hline
% \end{tabular}
%     \caption{P-values for Wilcoxon rank sum comparing weekly time spent values, rounded to two decimal places. Values over 0.05 indicate insignificant differences. }
%     \label{tab:weekly_wilco}
% \end{table}

Figure \ref{fig:weekly_time} shows the range of times in hours spent by students in the BSc CS and each week of the AI pop song content. The boxes contain 50\% of the data, and the lines in the boxes show the median. The whiskers show where 1.5 x the interquartile range falls, which for normal distributions contains 99\% of the dataset. Therefore anything outside the whiskers is considered an outlier. Students spend between 0.5 and 6 hours studying per week per course on the BSc CS and a similar amount on the AI course, except for the MusicVAE week, where students appear to spend considerably more time. We ran Wilcoxon rank-sum tests between the six distributions to check if the variation was significant. We chose this non-parametric test as we do not want to assume a normal distribution of the values. We found that the distribution of time spent on the AI content was significantly different from the BSc CS except for week 4 (Diffsinger). Generally, time spent in AI weeks appeared slightly lower than BSc CS, but week 3 (MusicVAE) was higher. 

Figure \ref{fig:lab_time} shows the range of times in minutes spent on the labs in the AI course and in the BSc CS as a whole. BSc students generally spend 5-35 minutes on labs, but many outliers are probably indicative of the wide variety of labs in the degree. The Wilcoxon rank-sum test did not find any significant differences between the observed time spent on the AI labs and the labs across the BSc CS. The larger-looking time band for AI week 5 is not significantly different from the other weeks. 

\subsubsection{Discussion of quantitative data}
Students spend less time per week on the AI course than on the degree, except for week 3, where they spent significantly more time. This metric only considers time spent on the platform indicated by course item hits, so it does not take into account activity outside of the platform and ignores downloading and watching videos later. 

Possible reasons for the lower time spent per week vs the whole BSc are that we placed the content at the end of the course, and it was credit-assessed with an exam as opposed to coursework that was more directly linked to the activities. We have now moved the content earlier in the course, and the planned MOOC version will have a different assessment model since it will not have paid markers, only peer reviews. Interestingly, the time spent on the MusicVAE content is significantly higher than in other weeks. This week has a more structured lab worksheet with multiple scripts providing a range of experiments for students to engage with. Also, they can generate actual music they can hear this week. Concerning the time spent in labs, some students spent considerable time in the labs, whereas others spent much less. Whilst the results here are only partially gratifying, there is much high-quality, high-rated content on the degree to compete with. These metrics certainly give us a benchmark against which to evaluate changes we make to the course.

\subsection{Qualitative evaluation: experts workshop}

In this section, we describe a qualitative analysis based on a workshop we ran with a curated panel of experts. The main goal of the workshop was to find out how we can improve the existing AI course. Specifically, we aimed to address the following research questions: 

\begin{enumerate}
\item How can we improve the existing AI course for creative practices?  
\item What are the current limitations in the course?
\item How can we improve the course for a diverse Coursera audience?
\end{enumerate}

\subsubsection{Method}
We organised a three-hour workshop in November 2022 with a multidisciplinary group of people including a music composer, distance learning expert, user evaluation expert, and AI-creativity educators and practitioners. We take note of \cite{colton2015stakeholder} who talked about the importance of considering different stakeholders in assessing the interaction between AI software and creative practice. 
%to expose our current course design and open the floor to suggestions and potential improvements of the course for creative practices. The three-hour workshop took place in November 2022, with some attending in-person and some online. The 8? attendees included creative practitioners,  
We started the workshop with a presentation consisting of an overview of the online BSc CS, a brief description of the AI and creativity field, and finally the high-level aims of the workshop.
In the second part of the workshop, we presented our course design and opened up the floor to an immediate response from the workshop panel. %which started to engage in interesting conversations around AI, creativity, ethics, and suggested a plan to improve and evaluate the existing AI course. 
The workshop was fully recorded, together with a transcript.

% In light of the characteristics of the collected transcript data, as well as our emphasis on \textit{sense making}, \textit{understanding}, and \textit{giving someone a voice}, we relied on Braun and Clarke's Reflexive Thematic Analysis \cite{clark_2019} as opposed to other qualitative methodologies such as Grounded Theory \cite{willig_2008} or IPA \cite{larkin_2006}. The six-step process of Reflexive Thematic Analysis by Braun and Clarke (familiarizing yourself with the data, generating initial codes, generating initial themes, reviewing themes, defining and naming themes, and producing the report) was carefully executed using a semi-inductive approach (see supplementary material for a detailed description of the process).
To analyse the text transcript of the workshop we employed Braun and Clarke's Reflexive Thematic Analysis (RTA)\cite{clark_2019}. Other qualitative methodologies exist such as Grounded Theory\cite{willig_2008} and Interpretative Phenomenological Analysis\cite{larkin_2006} but we chose RTA due to its emphasis on \textit{sense making}, \textit{understanding}, and \textit{giving someone a voice} and its clear protocol. 
We carefully executed the six-step protocol of RTA (familiarizing yourself with the data, generating initial codes, generating initial themes, reviewing themes, defining and naming themes, and producing the report) using a semi-inductive approach\footnote{See the appendix for a detailed description of the process}.

\subsubsection{Results and analysis}

\begin{figure}
    \centering
    \includegraphics[width=0.5\textwidth]{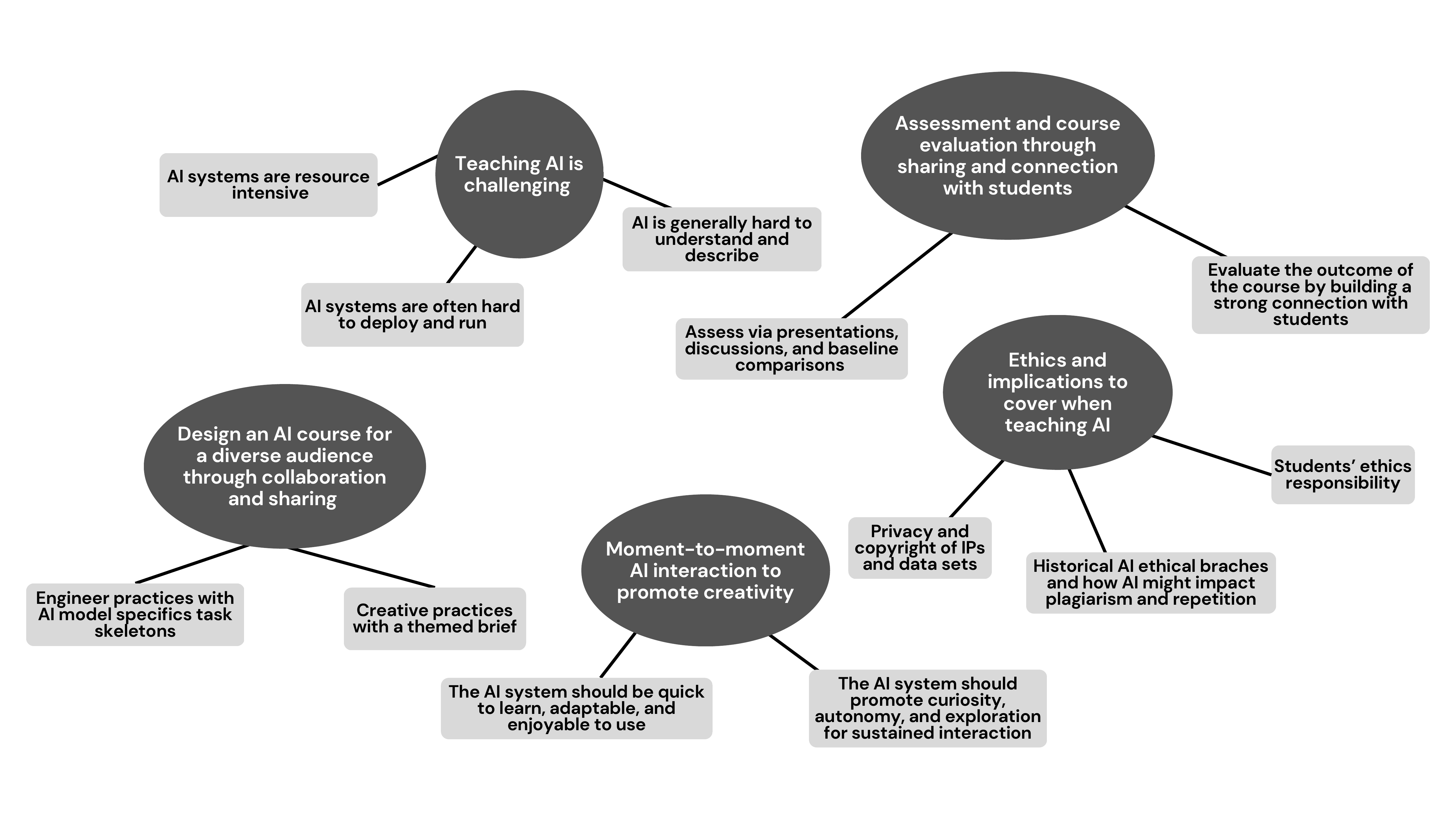}
    \caption{The final generated five themes and sub-themes from the Thematic Analysis process.}
    \label{fig:ta_themes}
\end{figure}

After carrying out the steps above, we identified five themes as shown in figure \ref{fig:ta_themes}. For each theme, we give a title, some sub-themes, some exemplary quotes and, concerning our main research goal, suggestions for future improvements of the course\addtocounter{footnote}{-1}\addtocounter{Hfootnote}{-1}\footnotemark.
%\footnote{see supplementary material for a detailed description of the }. 

\textit{Teaching AI is challenging:} The process of teaching AI can be really challenging. AI systems can be hard to describe and understand due to their complex structure and parameters (``GPT-3 has 175 billion parameters and uses all kinds of heavily iterated technology like."). Even if we take a valid pedagogical approach to deliver instructional material about AI, these systems are generally resource intensive and require a lot of skills and time to deploy and run (``You wouldn't be able to load the model into any hardware that any institution has in the UK, because it requires too much memory."). Suggestion: Whilst we have already taken account of this in the course design with the simpler GPT-2 model, we can consider linking in GPT-3 via a cloud service.
%consider the use of more powerful virtual labs to allow students the possibility to re-train the model with different data or provide them with instructions on how to train the model locally.

\textit{Moment-to-moment AI interaction to promote creativity:} A great element when it comes to creativity is exploration (``The creative concept is this idea of exploring this space"). AI systems should allow the user some autonomy and clearly communicate the possible interactions at any given moment (``Because they've chosen the task, they are motivated and they want to solve those problems."). These systems should be easy to learn, adaptable, and enjoyable to use in order to promote curiosity and sustained interaction (``About a sustained interaction that actually the AI tool is sufficiently good that you actually want to carry on using."). Suggestion: consider modifying the exercise worksheets to allow deeper student exploration of each of the components of the AI pop song generator.

\textit{Design an AI course for a diverse audience through collaboration and sharing:} We should take into consideration the idea of designing the course for a diverse audience. The industry may be happy with the way engineers approach AI but there seems to be a lack of individuals that can use AI in a creative way (``There aren't enough people that have a kind of an artistic sensitivity and understand the creative process and understand what it is to really explore and use AI to explore the creative content and production space."). The course should retain its technical content but at the same time, it should provide a themed creative brief for the less technical and more creative individuals (``Almost like an artistic or creative motivation for wanting you to show off what they can do. Yeah, this is the theme that I chose."). The diverse audience can come together via collaborative tasks and sharing of the obtained results (``So there's something presumably here about students understanding the different ways that they approach it in each other."). Suggestion: restructure the course so that individuals can approach it either creatively or in a more engineer-type way. We are planning to introduce role-playing within the course to tackle assignments from different perspectives.

\textit{Assessment and course evaluation through sharing and connection with students:} The course should provide a valid method to evaluate both the technical and the creative domains of AI. The produced work can be compared to a baseline project (``Maybe compared to baseline? So what's the simplest and stupidest thing that you can produce with this software"). Students should be encouraged to share their work via presentations and forum discussions (``You have to present your work or you have to put in a forum."). In terms of the course evaluation, the course leader can invite students' cohorts back to learn about their learning experiences (``Could ask for students or participants to kind of volunteer to stay in touch with you."). Suggestion: provide a valid method to assess both the technical and creative work. We can assess the produced work against a baseline project and give students plenty of opportunities to share their work and get feedback from other students.

\textit{Ethics and implications to cover when teaching AI:} When it comes to AI, ethics is a really important topic to consider, especially in a course when we are potentially introducing new people to this field. The course should cover the implications of using AI for creative practices: from IP and dataset copyrights to awareness of ethics breaches in the history of creativity (``Specific creative industries IP issues, right? Because um it's certainly something which impacts a lot of people in creative industries"). Part of the responsibility of the ethics should be in the interest of students (``We're teaching a course like this is their responsibility to revise at least some ethics in tandem with this"). Suggestion: introduce another week of content covering the ethics of using AI with concrete examples and possible implications.

\section{Concluding remarks and future work}

In the introduction, we enumerated three intended contributions. 
%
% We have delivered on these by providing details of a 
First, in outlining a new AI-Creativity course (now in its third run), we have provided a pedagogical, technical, and evaluation platform for other educators to use. 
%
% Future work will be focused on attracting a wider community of learners with different technical and creative backgrounds to interact state-of-the-art AI systems for creative content generation. 
%
%
Second, we have designed and applied a benchmarking evaluation method for the quantitative analysis of student engagement. 
Third, we demonstrated an orthogonal qualitative approach which engaged stakeholders in designing the course  
% as a reflexive thematic analysis' approach of a workshop 
and we have shown how it has revealed important themes along which we need to reflect, design, and test our AI-creativity materials.
%

%These distinct approaches will allow us to integrate into our ongoing design, important developments and themes in AI and related Creative workflows (in industry and in artistic creative practices) and to evaluate how future iterations of the engage deeply with AI-creativity content. 

%The fact we have used both approaches ensures that we can be certain that new developments in AI and related Creative workflows (in industry and in artistic creative practices) are reflected in our design and that our course design is sufficiently motivating to ensure students sustain their engagement over the length of the course. 

In future work, we will develop the course into a standalone MOOC with additional material relating to ethical and legal aspects of AI-creativity. 
 We will adapt our learning activities for more sustained student engagement through deeper exploratory activities motivated by using creative briefs.
We are excited by the MOOC format, as the opportunities will be far greater for iterated experimentation with different styles of assessment, pre- and post-testing of knowledge, and other factors such as evaluating the quality of musical output through peer review. 
%
%We look forward to future, ever-improving iterations of the course being taken by a far broader cohort as we are keen to ensure that students from a wide-range of disciplinary backgrounds and experiences can all access the creative potential of AI.
%

We are excited that the emergence of accessible AI-creativity tools and the field's cross-disciplinary nature can, with carefully designed learning materials, provide opportunities to include AI-creativity in wider areas of university and even school curricula. 

\bibliographystyle{named}
\bibliography{7_ijcai23}

\newpage

\appendix

\onecolumn

\begin{center}
    \huge{Give professionals a voice on an online course to teach collaborative, creative AI: A Thematic Analysis approach}
\end{center}

\vspace{1cm}

\section{Introduction and Overview}
This document acts as supplementary material for the main paper \textit{``The pop song generator: designing an online course to teach collaborative, creative AI."} More precisely, it offers a more detailed and comprehensive description of the qualitative analysis method adopted to evaluate the BSc in Computer Science AI course delivered using the Coursera learning platform.

The AI course is part of the curriculum of a fully certified online BSc Computer Science degree launched by Goldsmiths University and the University of London in early 2019. Through one of its case studies, the pop-song generator, the course aims to teach AI not just to practitioners in the science domain, but also to practitioners in the creative arts. During the five weeks, the length of the pop song case study, students work towards the generation of a music jingle combining a language model (GPT-2), a symbolic music model (Music-VAE), and a singer synthesizer (Diffsinger).

As part of our cross-evaluation process for the course, we organised a three-hour workshop with professionals in the AI field. The purpose of the workshop was to gather feedback from a multidisciplinary group of people around the following questions: How can we improve the existing AI course for creative practices? What are the current limitations? How can we improve the course for a diverse Coursera audience?

The fact that we wanted to give professionals a voice, the type of feedback obtained after the workshop, and the data collection process suggested the use of Thematic Analysis to carry out the qualitative evaluation. In order to follow a systematic and methodical process, we relied on Braun and Clarke's emended Thematic Analysis method, which they now refer to as Reflexive Thematic Analysis. Braun and Clarke provide an outline guide for conducting Reflexive Thematic Analysis made of six steps: familiarising yourself with your data, generating initial codes, generating initial themes, reviewing themes, defining and naming themes, and producing the report.

The six methodical Reflexive Thematic Analysis steps adopted with the workshop transcribed data revealed 35 initial codes which, by the end of the analysis process, boiled down to the following themes: ``Challenge of teaching AI", ``Design for different audiences", ``Promote creativity", ``Evaluation", and ``Ethics and implications."

\section{Presenting the AI Pop Song Case Study}

Following on approaches from Piaget and Henriksen, we designed the pop song case study to engage with real-world AI complexity in an exploratory manner. The case study is part of a course entitled ‘Artificial Intelligence’ and it is a final year undergraduate course offered as part of a certified BSc in Computer Science degree on the Coursera learning platform.

During the five-week course, which is the length of the pop song case study, students explore the creative domain, techniques, and components of using AI to generate a pop song. Students learn concepts such as auto-regression, latent spaces, and model orchestration as they explore the individual AI components that construct the pop song generator. More precisely, students engage with models such as GPT-2 for the lyrics generation, Music-VAE for the music generation, and Diffsinger for the vocals generation. This constructionist approach allows students to understand the individual AI components of the pop generator system and, at the same time, it provides them with some degree of exploration at each stage of the learning process.

As well as lectures, readings, quizzes, and discussion prompts, the case study is supported by the use of Coursera Virtual Labs. These environments allow learners to seamlessly work on projects and assignments in a browser without any local setup or software downloads. This is actually a great feature of the course which helps to overcome some of the challenges of teaching AI. Students not being able to afford powerful machines to train the systems, machines' limitations, and the complex and time-consuming art of setting up a local environment are all part of the challenges in teaching AI. For the AI pop song case study, we set up the Coursera Labs to run Visual Studio Code, Python, and Node with all the required packages to build and explore the pop song AI system. All students have to do is launch a Coursera Lab, follow the written instructions for a particular task, and they are all set to use and explore a particular component of the pop song generator.

By the end of the case study, students should have a good understanding of the individual components that make the AI pop song system and, moreover, they should have a complete AI system that can generate a pop song. At this stage, students are encouraged to share their generated songs with other students using discussion prompts, forums, and peer review assignments. They have full access to the code for the components of the system in the Coursera Labs for the rest of the AI course and can easily edit and experiment more with it and, why not, generate more songs.

\newpage

\section{Organising a Workshop with AI Professionals}
As part of our cross-evaluation process for the course, we wanted to gather feedback from both students in our course and professionals in the field of AI, music, and education. While we extracted quantitative analytics to evaluate the students' experiences within the AI pop song case study, our focus was also to hear the various perspectives from professionals on how we could improve the course for creative practices. Particularly, responses around the following questions: How can we improve the existing AI course for creative practices? What are the current limitations in the course? How can we improve the course for a diverse Coursera audience?

This led us to organise a workshop with a multidisciplinary group of people to expose our current course design and open the floor to suggestions and potential improvements of the course for creative practices. The three-hour workshop took place on 24 November 2022 in our filming studio at Goldsmiths, University of London. This is the same studio we used to film all the BSc Computer Science online lectures. It is equipped with hardware and custom software which makes it possible to create a live edit of a video with multiple camera shots and a green screen. Professionals in the field of AI, education, music, and ed-tech joined the workshop either in person or online via Microsoft Teams for a total of 10 attendees.

The workshop started with a brief history of Goldsmiths computing and a general overview of the online BSc in Computer Science to date. It then covered a brief history of AI and creativity and introduced the high-level aim of the workshop:

\begin{itemize}

    \item{Consider the current state of AI (especially regarding the last 10 years of massive investment and progress/AI-creativity systems) and the challenges we have when engaging a diverse group of people in learning how to use it at various levels}
    \item {Outline what we have done in terms of our innovations with the AI course}
    \item {Capture the immediate reactions to what we are trying to do}
    \item {Think through ways in which we might evidence pedagogical benefit/impact for students}
\end{itemize}

In the second part of the workshop, we presented our AI pop song case study and outlined aspects of the course design, the challenges of teaching AI for creative practices, the technology and pedagogy approach used in the course, and the course analytics from the first two sessions. The presentation was then followed by an immediate response from the workshop panel which opened up the floor to some interesting conversations around AI, creativity, ethics, and suggested a plan to improve and evaluate the existing AI course. The workshop was recorded in full, together with a transcript, so that we could later analyse all the responses and accurately take into account all ideas discussed and action points required.

\section{Selecting the Appropriate Method: Reflexive Thematic Analysis}

The three-hour workshop was very successful and well-appreciated by all the participants. The fact that a multidisciplinary group of people was involved produced interesting and strong conversations revolving around educational, ethical, and creative aspects of teaching AI.

In light of the characteristics of the collected transcript data, as well as our emphasis on \textit{sense making}, \textit{understanding}, and \textit{giving someone a voice}, it was suggested the use of qualitative analysis to evaluate the workshop outcome. We really aimed to put participants at ease and capture useful information about their opinion on our questions of interest: How can we improve the existing AI course for creative practices? What are the current limitations? How can we improve the course for a diverse Coursera audience? 

At first glance, Grounded Theory seemed an appropriate qualitative methodology path to discover possible theories on how to improve the existing AI course. In essence, Grounded Theory is both the process of category identification
and integration (as a method) and its product (as a theory), concerned with identifying and constructing theory from data. Unfortunately, the primary concern was with the nature and scale of the qualitative study. First of all, Grounded Theory operates with theoretical sensitivity, where the researcher interacts with the data in an iterative manner. Such a type of interaction, which usually requires the researcher to analyse data as it is collected, was not compatible with the nature of this qualitative study as the data collection process ended prior to beginning the analysis. Second, the relatively small scale and time frame of the qualitative study posed a risk of incomplete adoption of Grounded Theory as a methodology.

IPA (Interpretative Phenomenological Analyses) was also reviewed as a possible methodological approach but it was soon ruled out due to incompatibility with the nature of the study and the data collection analysis. IPA's concern is with exploring people's lived experiences and the meaning people attach to those experiences. At the heart of this perspective lies a clearly declared phenomenological emphasis on the experiential claims and concerns of the people taking part in the study as discussed by Larkin, Watts, and Clifton. The purpose of the workshop did not really seek to obtain information in regard to lived experiences; none of the participants had seen the course prior to attending the workshop. Furthermore, IPA requires each participant’s response to be examined in isolation to produce rich personal experience narratives, something which was rather complex with the transcript obtained at the end of the workshop.

TA (Thematic Analysis), differently from other methods such as Grounded Theory or IPA, does not prescribe methods of data collection, theoretical positions, epistemological or ontological frameworks. It only provides a method for identifying, analysing, and reporting patterns in qualitative data. Prior to Braun and Clarke publication of a clear set of procedures to conduct TA, which has also helped in recognising the research approach as a distinctive method, its adaptable characteristics had generated disbelief around the validity of such a qualitative practice.

In order to facilitate better TA practice and to clarify conceptual mismatches and confusions seen in published papers since 2006, Braun and Clarke revised their initial TA approach in a recent publication. The authors' amended TA method, which they now refer to as Reflexive Thematic Analysis, is best described as theoretically flexible only as a generic method; specific iterations of TA encode particular paradigmatic and epistemological assumptions about meaningful knowledge production and thus their theoretical flexibility is more or less constrained compared to the TA approach described in 2006. 

This recent vision of Reflexive Thematic Analysis, which takes into account the researcher's subjectivity and reflexivity, was adopted to analyse the transcript obtained from the workshop. The six-step process of Reflexive Thematic Analysis by Braun and Clarke (familiarizing yourself with the data, generating initial codes, generating initial themes, reviewing themes, defining and naming themes, and producing the report) was carefully executed using a semi-inductive approach. \textit{Semi} in the sense that we sort of forced the code generation process around our questions of interest. \textit{Inductive} in the sense that we let the data speak without trying to fit the data into pre-existing theory or framework. Themes were generated at a semantic level, that is they were identified within the explicit or surface meanings of the data and not beyond what a participant has said.

\section{Engaging with the step-by-step TA process}
The analysis was conducted manually without the help of any smart third-party tool to identify codes and themes. We only used Taguette, a free and open-source qualitative data analysis tool, to group similar codes together and store data extracts. We did not expose the data externally as we run the tool locally on our computers to preserve confidentiality.

\subsection{Familiarizing yourself with the data}
We collected the data through the interactive mean of running a workshop. Therefore, it is fair to say that we began the Thematic Analysis process with some prior knowledge of the data. 

Working with verbal data, we relied on the automatically transcribed document produced during the workshop. Microsoft Teams has in fact the option to automatically record and transcribe anything that has been said in a virtual meeting. At this point, we converted the transcribed document into a PDF, loaded it onto Taguette, and began with the first step of Thematic Analysis: familiarizing yourself with the data.

Following Braun and Clark's suggestions, it is ideal to read through the entire data set at least once before you begin your coding. During this phase, we read the full transcribed document on Taguette and started taking notes. All the conversations in that document were anonymous and we analysed the data without associating a particular comment with the participants. This process generated initial ideas for coding that became useful in subsequent phases of the analysis.

Here is a list of preliminary notes generated during this phase:

\begin{itemize}

\item{Think about that moment-to-moment interaction with AI rather than AI coming along as a collaborative tool}
\item{There is no interest in replacing anybody with AI}
\item{Importance of ethics on our undergraduate and postgraduate offerings when it comes to AI}
\item{Teaching state-of-the-art AI for creative practices is hard (the systems are challenging, sometimes the systems are not even explained in the books or are difficult to understand, systems are resource intensive, systems are hard to deploy and run)}
\item{There are not many people that have a kind of artistic sensitivity and understand what it is to really explore and use AI for creative content (the need for courses like this to support creative practices)}
\item{The need for AI systems to be able to adapt creatively to various scenarios and not become redundant very quickly. Usability is another important thread. It is about sustained interaction}
\item{Students who explore tend to get higher grades. The need to design a course that encourages students to explore}
\item{What are the strategies that we use to support the diversity of students? (e.g. engineers vs creative musicians)}
\item{Give students a starter pack. Structure the tasks around a series of deliverables (engineering part), let students explore (creative part), and have them finally come together and share their results at various points in the course}
\item{How can we judge if creativity is good enough in this course? Give students a brief or a theme on what they should explore with the AI system. Creative activities are not as well specified as the technical ones}
\item{Dealing with plagiarism and copyright with collaborative AI systems}
\item{Include a section at the end of the course to encourage students to showcase their work (live events, presentations)}
\end{itemize}

Once we had a list of initial ideas, we read the full data transcript one more time to see if we missed anything important at a surface level. As you can notice from the above list of ideas, the familiarisation process highlighted interesting points about improvements, limitations, and creative suggestions for our existing AI course. The captured ideas spawn compelling comments around challenges, ethics, creative practices, and evaluation methods to take into consideration when designing an AI course for a diverse audience. 

With this preliminary investigation, we were ready to move to the next step of the Thematic Analysis process: generating initial codes.

\subsection{Generating initial codes}

Right after generating an initial list of ideas in the data, we moved into the process of generating initial codes for the transcribed document. We approached the coding phase with specific questions in mind that we wanted to code around. Despite coding to identify  particular features  of  the  data set, we gave full and equal attention to each data item.

The coding phase was manually executed using Taguette. We worked systematically through the entire transcribed document, identified the codes, and associated them with data extracts that demonstrated that code. Following this process, all the data extracts that matched a particular code were then collated together with a code label in Taguette.

This phase of the Thematic Analysis produced an initial set of 35 codes with their respective data extracts. \autoref{tab_01_codes} and \autoref{tab_02_codes} show a list of all the generated codes with one example of the associated data extract for each of the codes.

\begin{table}[H]
\centering
\footnotesize
\caption{Generated codes with an example data extract (codes 1-18)} %title of the table
\rowcolors{2}{white}{gray!18}
\begin{tabular}{p{3cm} p{8cm}}% creating two columns
\hline %inserting double-line
\textbf{Code} & \textbf{Data extract}\\ [0.3cm]
\hline % inserts single-line
AI-interaction & The ongoing moment-to-moment interaction with AI rather than AI coming along. \\
Adaptability & I mean there are more complicated ways that people could use these systems in different contexts. \\
Autonomy & Because they've chosen the task, they are motivated and they want to solve those problems. \\
Challenging to describe & Systems are challenging to describe and understand. \\
Complex parameters & GT3 has 175 billion parameters and uses all kinds of heavily iterated technology like. \\
Creative interest & If you can motivate them by being creatively driven by something that they're interested in, then they will do more than if you just give them the basic engineering style kind of learning. \\
Curiosity & Set them a task which you know they will be led by the curiosity. \\
Exploration & If you can encourage everybody to do more exploring with the way you teach, then everybody will be sort of acting like a good student if you like. \\
Hard to deploy & The systems are hard to deploy and run. \\
IP issues & Specific Creative industries IP issues, right? Because um it's certainly something which impacts a lot of people in creative industries. \\
Lack of creative AI individuals & There aren't enough people that have a kind of artistic sensitivity and understand the creative process and understand what it is to really explore and use AI to explore and creative content, production space. \\
Quick learnability & It's hard and so, but it's especially if someone's coming in cold and you want to just get them up to speed. \\
Redundancy & If you can only barely run them once, you know it's. It's not actually that helpful. \\
Repetition & The likelihood of it coming up or simply repeating another. \\
Resource intensive & You wouldn't be able to load the model into any hardware that any institution has in the UK, because it requires too much
memory. \\
Share your work & Actually, you know what you want to do is go out there, find a training set and put it into this thing, and then get some results and share it in the forum. \\
Sustained interaction & About a sustained interaction
that actually the AI tool is sufficiently good that you actually want to carry on using it. \\
Task skeleton & So, if you know that some of your cohort struggled to get started, you give them a starter and then structurally then you can kind of structure the task around a series of deliverables which they know they have to share specific points. \\
\hline % inserts single-line
\end{tabular}
\label{tab_01_codes}
\end{table}

\begin{table}[H]
\centering
\footnotesize
\caption{Generated codes with an example data extract (codes 19-35)} %title of the table
\rowcolors{2}{white}{gray!18}
\begin{tabular}{p{3cm} p{8cm}}% creating two columns
\hline %inserting double-line
\textbf{Code} & \textbf{Data extract}\\ [0.3cm]
\hline % inserts single-line
Technical content & Regression is basically eating your own dog food or something like that. So, it's kind of feeding the output of the system back into the input, so it allows you to continually generate stuff. So that's the technical concept. \\
Usable and enjoyable & The best bit is when you get someone to use it and they enjoy using it. \\
Audience collaboration & So, there's something presumably here about students understanding the different ways that they approach it in each other. \\
Baseline comparison & Maybe compared to baseline? So, what's the simplest and stupidest thing that you can you that you can produce with this software. \\
Case law & So, there is existing case law that could support an, you know, an existing course. \\
Cohorts & With cohorts like inviting cohort back. \\
Connections with students & Could ask for students or participants to kind of volunteer to stay in touch with you. \\
Creative brief & Almost like an artistic or creative motivation for wanting you to show off what they can do. Yeah, this is the theme that I chose. \\
Ethics information & I think the inclusion of ethics to some extent needs to be there. \\
Evaluation questions & Doing a quick assessment of how expressive is this, how it technically well done? Is this? How much does it speak to you? So quick? \\
History of creativity & Yeah, and also fitting it into the kind of artistic history. \\
Lyrics & So, there are multiple outcomes, so the first one is the lyrics. \\
Mixed feedback & If you have some anonymous feedback and some not anonymous feedback, you know maybe that provides some opportunities. \\
Presentation & Would gain value from so you know, cause some kind of presentations or some kind of event, or some sort of activity. \\
Students’ ethics responsibility & We're teaching a course like this is their responsibility to revise at least some ethics in tandem with this. \\
Students impact & Way in which they operate in the future, thinking about whether you can kind of catch up with people six months. \\
Track students & Crawl people or the yeah consequences of the course. \\
\hline % inserts single-line
\end{tabular}
\label{tab_02_codes}
\end{table}

At the end of the coding process, we revisited the transcribed document once more to check that we were happy with the generated codes and that we did not miss any obvious and important information. The 35 initial codes were generated at a semantic level, that is they were identified within the explicit or surface meanings of the data and not beyond what a participant has said.

At the end of this phase, we were satisfied with the generated codes and we were ready to move to the next step of the Thematic Analysis: generating initial themes.

\newpage

\subsection{Generating initial themes}

At this stage, all data were initially coded and collated, and we had a long list of 35 different codes, and text extracts, that we identified across the transcribed document. We could therefore move into the generating initial themes phase of the analysis.

This phase, saw us refocusing on the analysis at the broader level of themes, rather than codes, and involved sorting  the  different  codes  into potential themes, and collating all the relevant coded data extracts  within the identified themes. In other words, we started to analyse our codes and considered how different codes would combine under a single overarching title/theme.

We started to analyse all the code extracts for individual codes using the Taguette software and noted down on paper potential overarching titles to house closely related codes. It was really helpful at this stage to use visual representations to help us sort the different codes into themes. We decided to use a combination of tables and mind maps to visually represent the outcome of this phase of the analysis.

Here is a series of tables that represent the generated preliminary themes:
\newline

\color{CadetBlue}
\begin{center}
\large{\textbf{Challenge of teaching AI}}
\end{center}
\color{black}
\normalsize
\begin{table}[H]
\centering
\begin{tabular}{ | m{2.5cm} | m{2.5cm}| m{2.5cm} | m{2.5cm} | } 
  \hline
  \cellcolor{blue!10} Challenging \newline to describe & \cellcolor{blue!10} Complex \newline parameters & \cellcolor{blue!10} Hard to deploy & \cellcolor{blue!10} \cellcolor{blue!10} Resource \newline intensive \\ 
  \hline
\end{tabular}
\end{table}

\color{Orange!80}
\begin{center}
 \large{\textbf{Promote creativity}}
\end{center}
\color{black}
\normalsize
\begin{table}[H]
\centering
\begin{tabular}{ | m{2.5cm} | m{2.5cm}| m{2.5cm} | m{2.5cm} | } 
  \hline
  \cellcolor{Apricot!30} AI-interaction & \cellcolor{Apricot!30} Adaptability & \cellcolor{Apricot!30} Autonomy &  \cellcolor{Apricot!30} Creative \newline interest \\ 
  \hline
   \cellcolor{Apricot!30} Curiosity & \cellcolor{Apricot!30} Exploration & \cellcolor{Apricot!30} Quick \newline learnability &  \cellcolor{Apricot!30} Redundancy \\ 
   \hline
    \cellcolor{Apricot!30} Sustained \newline interaction & \cellcolor{Apricot!30} Usable and \newline enjoyable & \cellcolor{Apricot!30} & \cellcolor{Apricot!30} \\
    \hline
\end{tabular}
\end{table}

\color{LimeGreen}
\begin{center}
\large{\textbf{Design for different audiences}}
\end{center}
\color{black}
\normalsize
\begin{table}[H]
\centering
\begin{tabular}{ | m{2.5cm} | m{2.5cm}| m{2.5cm} | m{2.5cm} | } 
  \hline
   \cellcolor{LimeGreen!20} Lack of creative AI individuals & \cellcolor{LimeGreen!20} Task skeleton & \cellcolor{LimeGreen!20} Technical \newline content &  \cellcolor{LimeGreen!20} Audience \newline collaboration \\ 
   \hline
    \cellcolor{LimeGreen!20}  Creative brief  & \cellcolor{LimeGreen!20} Lyrics & \cellcolor{LimeGreen!20} & \cellcolor{LimeGreen!20}  \\
    \hline
\end{tabular}
\end{table}

\color{Dandelion!80}
\begin{center}
\large{\textbf{Work evaluation}}
\end{center}
\color{black}
\normalsize
\begin{table}[H]
\centering
\begin{tabular}{ | m{2.5cm} | m{2.5cm}| m{2.5cm} | m{2.5cm} | } 
  \hline
  \cellcolor{Yellow!20} Share your work & \cellcolor{Yellow!20} Baseline \newline comparison & \cellcolor{Yellow!20} Cohorts &  \cellcolor{Yellow!20} Connections with students \\ 
  \hline
   \cellcolor{Yellow!20} Evaluation questions & \cellcolor{Yellow!20} Mixed feedback & \cellcolor{Yellow!20} Presentation &  \cellcolor{Yellow!20} Students impact \\ 
   \hline
    \cellcolor{Yellow!20} Track students & \cellcolor{Yellow!20} & 
    \cellcolor{Yellow!20} & 
    \cellcolor{Yellow!20} \\
    \hline
\end{tabular}
\end{table}

\color{Mulberry}
\begin{center}
\large{\textbf{Ethics and implications}}
\end{center}
\color{black}
\normalsize
\begin{table}[H]
\centering
\begin{tabular}{ | m{2.5cm} | m{2.5cm}| m{2.5cm} | m{2.5cm} | } 
  \hline
   \cellcolor{Thistle!30} IP issues & \cellcolor{Thistle!30} Repetition & \cellcolor{Thistle!30} Case law &  \cellcolor{Thistle!30} Ethics \newline information \\ 
   \hline
    \cellcolor{Thistle!30}  History of \newline creativity  & \cellcolor{Thistle!30} Students’ ethics responsibility & \cellcolor{Thistle!30} & \cellcolor{Thistle!30}  \\
    \hline
\end{tabular}
\end{table}

\newpage

Here is the mind map representation of the generated preliminary themes:

\begin{figure}[h]
\centering
\fbox{\includegraphics[width=\textwidth]{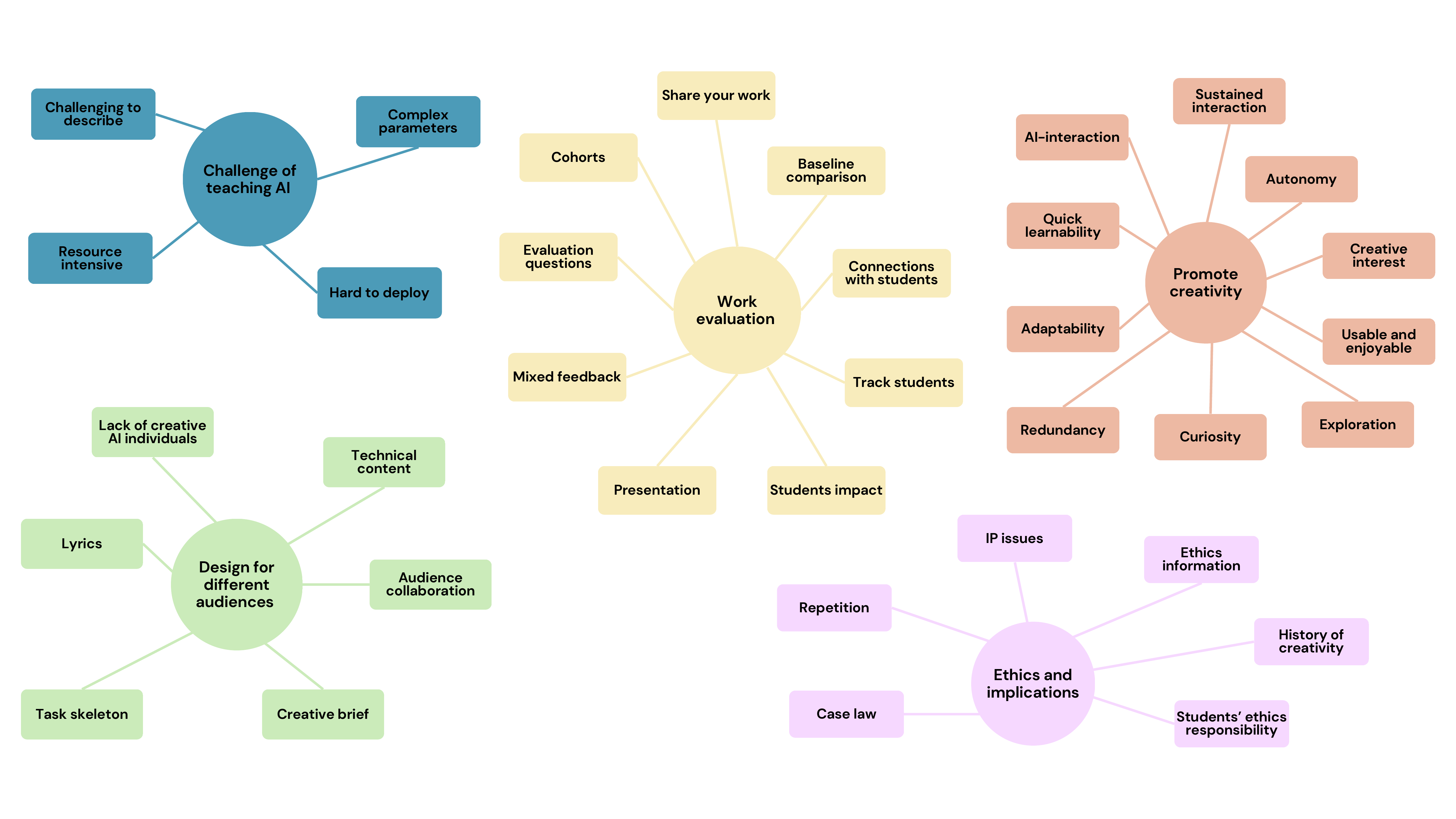}}
\caption{Mind map of the five generated themes}
 \end{figure}

This phase of the Thematic Analysis process resulted in a collection of five candidate themes and all the text extracts that were coded in relation to them. The five generated themes were: challenge of teaching AI, promote creativity, design for different audiences, work evaluation, and ethics and implications. We ended this phase with a brief and broad description of the five generated themes with examples of text extracts.  
\newline
\newline
\newline
\newline
\textbf{Challenge of teaching AI:} The process of teaching AI can be really challenging. AI systems can be hard to describe and understand due to their complex structure and parameters (``GPT-3 has 175 billion parameters and uses all kinds of heavily iterated technology like."). Even if we take a valid pedagogical approach to deliver instructional material about AI, these systems are generally resource intensive and require a lot of skills and time to deploy and run (``You wouldn't be able to load the model into any hardware that any institution has in the UK, because it requires too much memory.").
\newline
\newline
\textbf{Promote creativity:} A great element when it comes to creativity is exploration (``The creative concept is this idea of exploring this space"). AI systems should allow the user some autonomy and clearly communicate the possible interactions at any given moment (``Because they've chosen the task, they are motivated and they want to solve those problems."). These systems should be easy to learn, adaptable, and enjoyable to use in order to promote curiosity and sustained interaction (``About a sustained interaction that actually the AI tool is sufficiently good that you actually want to carry on using.").
\newline
\newline
\textbf{Design for different audiences:} We should take into consideration the idea of designing the course for a diverse audience. The industry may be happy with the way engineers approach AI but there seems to be a lack of individuals that can use AI in a creative way (``There aren't enough people that have a kind of an artistic sensitivity and understand the creative process and understand what it is to really explore and use AI to explore the creative content and production space."). The course should retain its technical content but at the same time it should provide a themed creative brief for the less technical and more creative individuals (``Almost like an artistic or creative motivation for wanting you to show off what they can do. Yeah, this is the theme that I chose."). The diverse audience can come together via collaborative tasks and sharing of the obtained results (``So there's something presumably here about students understanding the different ways that they approach it in each other.").
\newline
\newline
\textbf{Work evaluation:} The course should provide a valid method to evaluate both the technical and the creative domains of AI. The produced work can be compared to a baseline project (``Maybe compared to baseline? So what's the simplest and stupidest thing that you can produce with this software"). Students should be encouraged to share their work via presentations and forum discussions (``You have to present your work or you have to put in a forum."). In terms of the course evaluation, the course leader can invite students' cohorts back to learn about their learning experiences (``Could ask for students or participants to kind of volunteer to stay in touch with you.").
\newline
\newline
\newline
\newline
\textbf{Ethics and implications:} Ethics are a really hot topic in the field of AI. The course should cover the implications of using AI for creative practices: from IP and dataset copyrights to awareness of ethics breaches in the history of creativity (``Specific creative industries IP issues, right? Because um it's certainly something which impacts a lot of people in creative industries"). Part of the responsibility of the ethics should be in the interest of students (``We're teaching a course like this is their responsibility to revise at least some ethics in tandem with this").

With a collection of five
candidate themes, all the text extracts that were coded in relation to them, and a broad description for each of the themes, we were ready to move into the next phase of the Thematic Analysis process: reviewing the themes.

\subsection{Reviewing the themes}
We started this phase of the Thematic Analysis process with a set of candidate themes and it was time for us to refine those. The refinement procedure consists in analysing the themes and code extracts in depth. Some themes may collapse  into  each  other while others might need to be  broken down into separate themes or sub-themes. Clear guidance from Braun and Clark suggests that data within themes should cohere together  meaningfully,  while  there  should be  clear  and identifiable distinctions between themes.

We approached this phase with two levels of review. In the first level, we used Taguette to read once more all the collated extracts for each theme and considered whether they appeared to  form a coherent pattern. In the second level, we used Taguette to analyse the themes in relation to the entire  data set. We considered the validity of individual themes in relation to the data set, but also whether the produced thematic map reflected the meaning of the data set as a whole. Similarly to the previous step, we decided to use a
combination of tables and mind maps to visually represent the outcome of this phase of the analysis. 

Here is a series of tables that represent the outcome of the first level of the themes review with a description of the changes that were made to each of the themes during the review process. 
\newline

\color{CadetBlue}
\begin{center}
    \large{\textbf{Challenge of teaching AI}}
\end{center}

\color{black}
\normalsize
\begin{table}[H]
\centering
\begin{tabular}{ | m{5.5cm} | m{5.5cm}|} 
  \hline
  \multicolumn{2}{|c|} {\cellcolor{blue!10} Resource intensive} \\
  \hline
   \multicolumn{2}{|c|} {\cellcolor{blue!10} 
      Hard to deploy and run}  \\  
  \hline
  \multicolumn{2}{|c|}{\cellcolor{blue!10} Hard to understand} \\ 
  \hline
  Complex parameters & Difficult to describe \\
  \hline
\end{tabular}

\end{table}

The theme `Challenge of teaching AI' did not change much from the previous iteration. We only restructured the theme to have three sub-themes: resource intensive, hard to deploy and run, and hard to understand. The latter sub-theme was further supported by two codes: complex parameters and difficult to describe.

\color{Orange!80}
\begin{center}
    \large{\textbf{Promote creativity}}
\end{center}

\color{black}
\normalsize
\begin{table}[H]
\centering
\begin{tabular}{ | m{2.8cm} | m{2.5cm} | m{2.5cm} |  m{2.5cm} |}  
  \hline
  \multicolumn{4}{|c|}{\cellcolor{Apricot!30} Moment-to-moment AI interaction}   \\ 
  \hline
  \multicolumn{4}{|c|}{\cellcolor{Apricot!30} AI system characteristics}   \\  
  \hline
  Quick \newline learnability  & Adaptability  & Usable and \newline enjoyable & Redundancy \\
   \hline
  \multicolumn{4}{|c|}{\cellcolor{Apricot!30} AI and student interaction}   \\  
  \hline
  Curiosity  & Exploration  & Autonomy & Sustained \newline interaction \\
  \hline
\end{tabular}
\end{table}

The theme `Promote creativity' changed quite significantly in terms of its structure. We added three sub-themes: moment-to-moment AI interaction, AI system characteristics, and AI and student interaction. The AI system characteristics sub-theme highlighted interesting features of how to design an AI system to promote creativity. The sub-theme in question was further supported by the following codes: quick learnability, adaptability, usable and enjoyable, and redundancy. The AI and students interaction sub-theme, on the other hand, highlighted important aspects of the type of interaction that students should experience with an AI system to promote creativity. The sub-theme is further supported by the following codes: curiosity, exploration, autonomy, and sustained interaction.
\newline

\color{LimeGreen}
\begin{center}
\large{\textbf{Design for different audiences}}
\end{center}

\color{black}
\normalsize
\begin{table}[H]
\centering
\begin{tabular}{ | m{2cm} | m{2cm} | m{2cm} |  m{2cm} | m{2cm} |}  
  \hline
  \multicolumn{5}{|c|}{\cellcolor{LimeGreen!20} Audience  collaboration through sharing}   \\ 
  \hline
  \multicolumn{2}{|c|}{\cellcolor{LimeGreen!20} Engineers} &  \multicolumn{3}{c|}{\cellcolor{LimeGreen!20} Creative practices}  \\
  \hline
  Task \newline skeleton  & AI model \newline specifications & Creative brief & Theme & Lack of \newline creative AI individuals \\
  \hline
\end{tabular}
\end{table}

The `Design for different audiences' theme was restructured to mainly have one sub-theme: audience collaboration through sharing. The latter sub-theme was then additionally split into two sub-themes: engineers and creative practices. The engineers sub-theme highlighted interesting information on how to design practical learning activities for this audience and it was further supported by the following codes: task skeleton, and AI model specifications. The creative practices sub-theme highlighted similar useful information but for the creative audience. The sub-theme was further supported by the following codes: creative brief, theme, and lack of creative AI individuals.

\color{Dandelion!80}
\begin{center}
\large{\textbf{Evaluation}}
\end{center}

\color{black}
\normalsize
\begin{table}[H]
\centering
\begin{tabular}{ | m{2cm} | m{2cm} | m{2cm} |  m{2cm} | m{2cm} |}  
  \hline
  \multicolumn{5}{|c|}{\cellcolor{Yellow!20} Work evaluation}   \\ 
  \hline
  \multicolumn{3}{|c|}{\cellcolor{Yellow!20} Share your work} &  \multicolumn{2}{c|}{\cellcolor{Yellow!20} Evaluation methods}  \\
  \hline
  Presentation & Forum & External platform & Mixed \newline feedback & Baseline comparison \\
  \hline
   \multicolumn{5}{|c|}{\cellcolor{Yellow!20} Course evaluation}   \\ 
   \hline
   \multicolumn{5}{|c|}{\cellcolor{Yellow!20} Connection with students}   \\ 
   \hline
   Students \newline impact & Track \newline students &  \multicolumn{3}{|c|}{Invite cohorts} \\
  \hline
\end{tabular}
\end{table}

The `Evaluation' theme was also restructured to capture the essence of the text extracts associated with that theme. We felt like changing the name of the theme from the previous iteration as the associated text extracts did not only reveal evaluation suggestions for the students' work but also for the course as a whole. We added two major sub-themes: work evaluation and course evaluation. The first sub-theme was split into two other sub-themes: share your work and evaluation methods. The second, instead, was given ad additional sub-theme layer that better described the parent sub-theme: connection with students. The share you work sub-theme was further supported by the following codes: presentations, forums, and external platforms. The evaluation methods sub-theme was further supported by the following codes: mixed feedback and baseline comparison. Finally, the connection with students sub-theme was further supported by the following codes: students impact, track students, and invite cohorts.
\newline

\color{Mulberry}
\begin{center}
    \large{\textbf{Ethics and implications}}
\end{center}

\color{black}
\normalsize
\begin{table}[H]
\centering
\begin{tabular}{ | m{2.4cm} | m{2.5cm} | m{2.5m} | m{2cm} |}  
  \hline
  \multicolumn{4}{|c|}{\cellcolor{Thistle!20} Privacy and copyright}   \\ 
  \hline
  \multicolumn{2}{|c|} {IP issues} &  \multicolumn{2}{c|}{Datasets}  \\
  \hline
  \multicolumn{4}{|c|}{\cellcolor{Thistle!20} Course ethics information}   \\ 
  \hline
  { Case law}  & { Repetition and plagiarism}  & \multicolumn{2}{|c|}{History of creativity and ethical breach}  \\
  \hline
   \multicolumn{4}{|c|}{\cellcolor{Thistle!20} Students’ ethics responsibility}   \\ 
  \hline
\end{tabular}
\end{table}

Finally, the `Ethics and implications' theme was restructured to have three sub-themes: privacy and copyright, course ethics information, and students' ethics responsibility. The privacy and copyright sub-theme was further supported by the following codes: IP issues and data sets. The course ethics information sub-theme was further supported by the following codes: case law, repetition and plagiarism, and history of creativity and ethical breach.

\newpage

In the second level of the review process, we used Taguette to analyse the themes in relation to the entire data set. We read through the transcript one more time and agreed that the generated themes fit the data extracts pretty well. Here is the final mind map representation of the generated themes during this phase of the analysis:

\begin{figure}[h]
\centering
\fbox{\includegraphics[width=\textwidth]{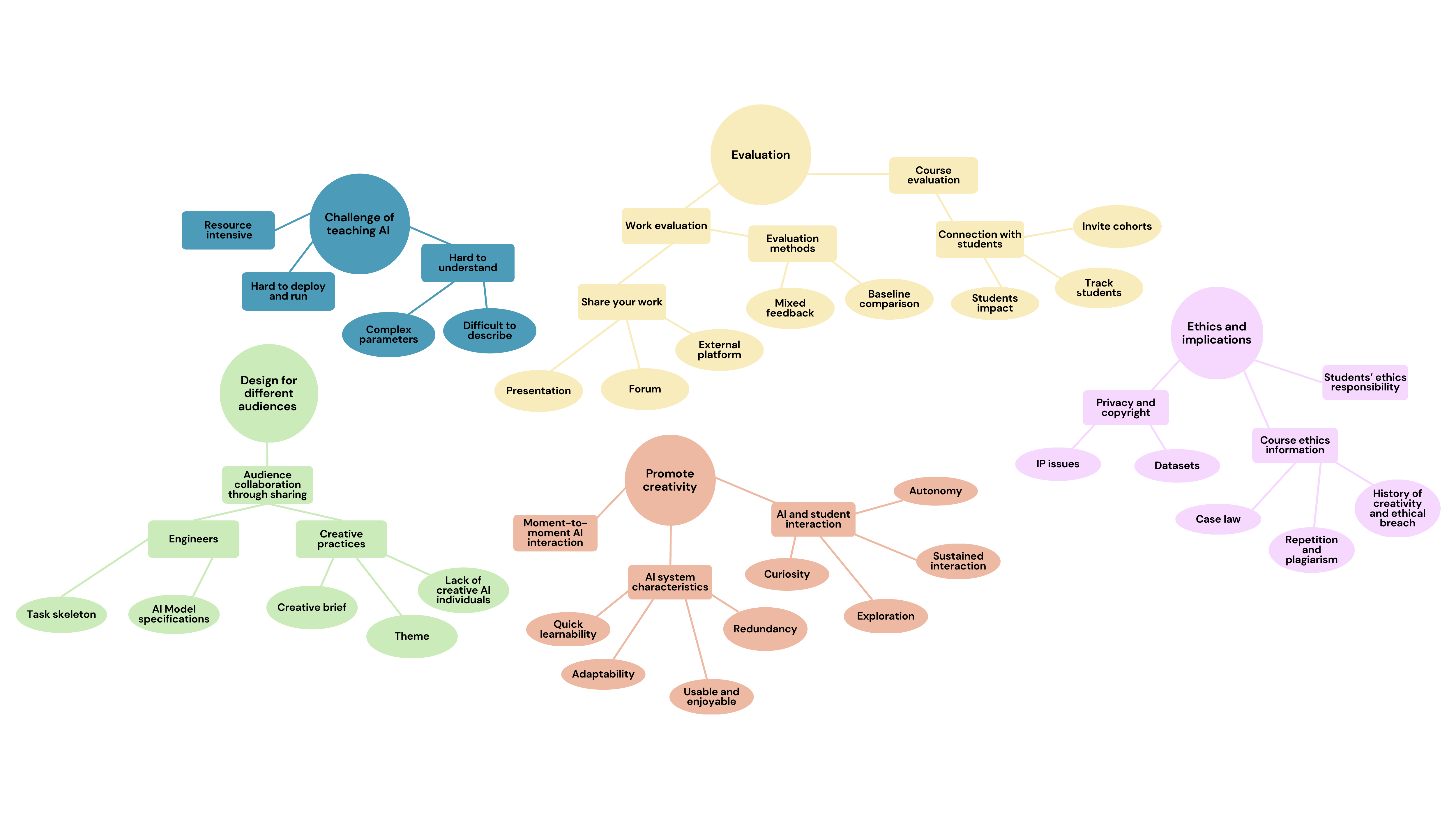}}
\caption{Mind map of the five generated themes with sub-themes}
 \end{figure}

 Now that we had a clearer structure of the potential themes, we were ready to move to the next, and almost final stage, of the Thematic Analysis process: defining and naming themes.

\subsection{Defining and naming themes}

Once we had a satisfactory thematic map of our data, it was time for us to move into this phase of the analysis: defining and naming themes. At this point, we spent time defining and further  refining the themes to identify their \textit{essence} and what aspects of the data each theme captured.
We did this by going back to the collated data extracts one more time and we then refined the thematic map so that it reflected as  accurately as possible the main point of interest for each of the generated themes. In other words, we simplified the thematic map so that each individual theme could be easily described in a  couple of sentences by just looking at the thematic map.

\newpage

Here is the final mind map representation of the generated themes during this phase of the analysis:

\begin{figure}[h]
\centering
\fbox{\includegraphics[width=\textwidth]{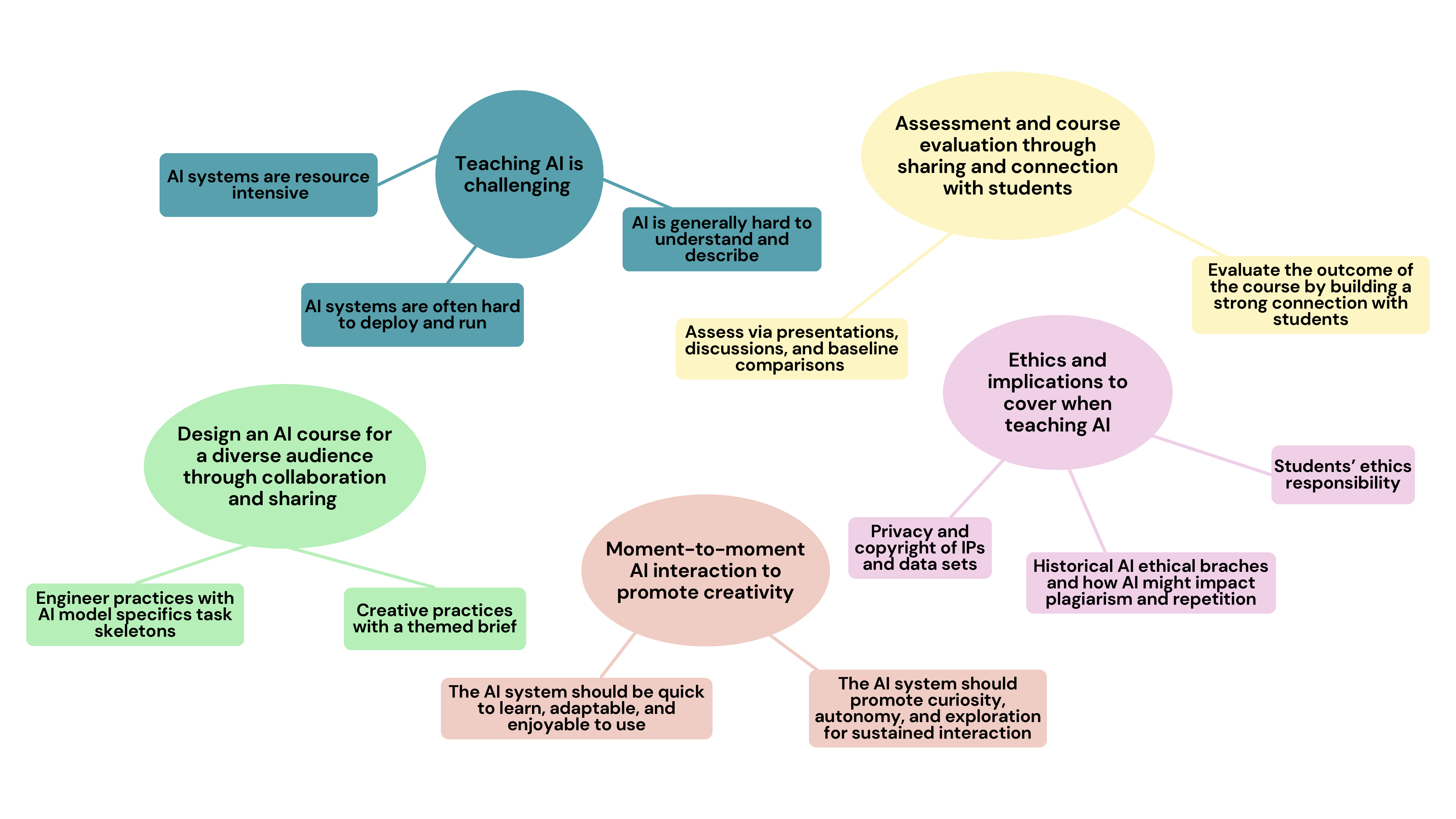}}
\caption{Mind map of the final five generated themes and sub-themes}
 \end{figure}

At this point, all the generated themes could be described in a few sentences and were coherent with the narrative of the research questions. Here is a series of tables representing the final themes with a short description that highlights the essence of each of the generated themes.

\color{CadetBlue}
\begin{center}
    \large{\textbf{Teaching AI is challenging }}
\end{center}

\color{black}
\normalsize

\begin{table}[H]
\centering
\begin{tabular}{ | m{11.5cm} |} 
  \hline
  \cellcolor{blue!10} AI systems are resource
 intensive \\
  \hline
   \cellcolor{blue!10} 
      AI systems are often hard to deploy and run  \\  
  \hline
  \cellcolor{blue!10} AI is generally hard to understand and describe \\ 
  \hline
\end{tabular}
\end{table}

Teaching AI is a challenging practice. AI material is generally difficult to understand and very tricky for instructors to present it in a concise, easy, and understandable way. AI systems have really complex parameters and functionalities and are usually resource intensive. This makes it quite difficult for these systems to be deployed successfully and run with no issues.

\color{Orange!80}
\begin{center}
    \large{\textbf{Moment-to-moment AI interaction to promote creativity }}
\end{center}

\color{black}
\normalsize
\begin{table}[H]
\centering
\begin{tabular}{ | m{11.5cm} |} 
  \hline
   {\cellcolor{Apricot!20} The AI system should be quick to learn, adaptable, and enjoyable to use} \\
  \hline
    {\cellcolor{Apricot!20} 
      The AI system should promote curiosity, autonomy, and exploration for sustained interaction}  \\  
  \hline
\end{tabular}
\end{table}

Creativity in AI can be promoted by analysing the moment-to-moment interaction between the user and the system. These systems should be quick to learn, easy to use, adaptable, and enjoyable to work with. The collaborative experience should promote curiosity, autonomy, and exploration in order to sustain that interaction.

\newpage

\color{LimeGreen}
\begin{center}
    \large{\textbf{Design an AI course for a diverse audience through collaboration and sharing}}
\end{center}

\color{black}
\normalsize
\begin{table}[H]
\centering
\begin{tabular}{ | m{11.5cm} |} 
  \hline
   {\cellcolor{LimeGreen!20} Engineer practices with AI model specifics task skeletons} \\
  \hline
    {\cellcolor{LimeGreen!20} 
     Creative practices with a themed brief}  \\
  \hline
\end{tabular}
\end{table}

Designing an AI course for a diverse audience involves all parties to collaborate and share their results. Engineers should be given skeleton tasks to explore the technical part of the AI system in more detail. On the other hand, creative practitioners should be provided with a themed brief where they can explore various possibilities with the AI system.

\color{Dandelion!80}
\begin{center}
    \large{\textbf{Assessment and course evaluation through sharing and connection with students}}
\end{center}

\color{black}
\normalsize
\begin{table}[H]
\centering
\begin{tabular}{ | m{11.5cm} |} 
  \hline
   {\cellcolor{Yellow!20} Assess via presentations, discussions, and baseline comparisons } \\
  \hline
    {\cellcolor{Yellow!20} 
     Evaluate the outcome of the course by building a strong connection with students}  \\
  \hline
\end{tabular}
\end{table}

Students' work and assessments should be evaluated using sharing activities like presentations and forum posts. The results should be then compared to a baseline project for technical and creative evaluation. There is a possibility to evaluate the course by building a strong relationship with the students and by inviting them back after they finished the course to share their experiences.

\color{Mulberry}
\begin{center}
    \large{\textbf{Ethics and implications to cover when teaching AI}}
\end{center}
\color{black}
\normalsize
\begin{table}[H]
\centering
\begin{tabular}{ | m{11.5cm} |} 
  \hline
   {\cellcolor{Thistle!20} Privacy and copyright of IPs and data sets } \\
  \hline
    {\cellcolor{Thistle!20} 
     Historical AI ethical braches and how AI might impact plagiarism and repetition}  \\
      \hline
    {\cellcolor{Thistle!20} 
    Students’ ethics responsibility}  \\
  \hline
\end{tabular}
\end{table}

Ethics and the implications of using AI systems should be included in the course. Students should be aware of the issues around the privacy and copyrights of IPs and data sets. Historical examples of AI ethical breaches may be one solution to make students more aware of the importance of ethics in AI. Students should be partially responsible for their use of AI systems and should make sensible choices when they collaborate with AI.

\section{Finishing up with the TA process and results}

We reached the final step of the Thematic Analysis, producing the report, with a set of fully worked-out themes. During this final stage of the analysis, we provided a detailed description of the generated themes and how they contributed to answer our initial research questions. 
\newline
\newline
\textbf{Teaching AI is challenging:} The process of teaching AI seems to be a really challenging task for instructors. Teaching a particular AI technique means understanding all the complex layers that form that particular system. In most cases, these systems have a really complex structure and a lot of parameters which can result in an endless permutation of possible uses to produce the desired result (``GPT-3 has 175 billion parameters and uses all kinds of heavily iterated technology like."). Instructors need to then summarise this complexity and present it  to their students in the most simplistic, yet elaborate, way for students to really understand the individual components that make these systems (``Systems are challenging to describe and understand"). Even if instructors take a valid pedagogical approach to deliver instructional AI material, the systems are generally resource intensive and require powerful machines to run them (``You wouldn't be able to load the model into any hardware that any institution has in the UK, because it requires too much memory."). Furthermore, AI systems can be really tricky to set up and run without spending a reasonable amount of time fiddling with the various technologies required to run them (``Next one is that systems are hard to deploy and run"). Our AI course uses virtual labs to remove the frustration of setting up the technologies needed to run the AI system and focuses on teaching all the individual components that make the AI pop singer generator. On the other hand, this theme highlights the fact that AI systems are generally resource intensive and require powerful machines to run them. This is a current limitation in our course as students are given a pre-trained model of the pop song generator due to the limited resources offered by the Coursera virtual labs.
\newline
\newline
\textbf{Moment-to-moment AI interaction to promote creativity:} The focus of promoting creativity in the field of AI should be on the moment-to-moment interaction between the user and the system rather than the overall collaboration between the two (``The ongoing moment-to-moment interaction rather than AI coming along"). AI systems designed for creative practices should have certain characteristics. First of all, they should be easy to learn; creative individuals with no engineering background would want to quickly learn how to use a particular AI system (``It's hard especially if someone is coming in cold and want to get up to speed."). Second, AI systems should be able to somehow adapt to various scenarios (``I mean there are more complicated ways that people could use these systems in different contexts") or at least produce diverse outcomes (``If you can only barely run them once you know it is not actually that helpful"). Finally, AI systems should be, to a certain extent, easy to use and enjoyable (``The best bit is when you get someone to use it and they enjoy using it."). The interaction between the user and the AI system is also extremely important to favour creativity. Exploration (``The creative concept is this idea of exploring this space"), autonomy (``Because they've chosen the task, they are motivated and they want to solve those problems."), and curiosity (``Set them a task which you know they will be led by the curiosity") are all fundamental aspects to maintain that sustained interaction between the user and the AI system. This theme highlights the need to design AI systems with certain characteristics in order to promote creativity and it really gave us useful insights on how to improve the existing AI course for creative practices.
\newline
\newline
\textbf{Design an AI course for a diverse audience through
collaboration and sharing:} The idea of wanting to design the course for a diverse audience sparked a lot of interesting conversations during the workshop.
The industry may be happy with the way engineers approach AI but there seems to be a lack of individuals that can use AI in a creative way (``There aren't enough people that have a kind of an artistic sensitivity and understand the creative process and understand what it is to really explore and use AI to explore the creative content and production space."). The diverse audience should approach the course as they wish, either creatively or in a more engineer-type way, and should come together via collaborative tasks and sharing of the obtained results at different stages in the course (``So there's something presumably here about students understanding the different ways that they approach it in each other."). Engineers should be oriented more towards technical tasks such as why the model is not good for certain things (``Interested in why the model does not work for certain things."), while creative individuals should be given a themed creative brief to work around (``Almost like an artistic or creative motivation for wanting you to show off what they can do. Yeah, this is the theme that I chose."). This theme highlights some of the aspects that we can adopt to design learning activities that are suitable for a diverse audience.
\newline
\newline
\textbf{Assessment and course evaluation through sharing and
connection with students:} The course should provide a valid method to evaluate both the technical and the creative domains of AI, as well as the course experience as a whole. Students should be encouraged to share their work via presentations and forum discussions (``You have to present your work or you have to put it in a forum."). They should also be encouraged to comment on each other's work with an appropriate feedback format depending on the type of learning activity (``If you have some anonymous feedback and some not anonymous feedback, you know maybe that provides some opportunities.") Assessing the creativity of a particular assignment can be really difficult without a proper comparison to a baseline project (``Maybe compared to baseline? So what's the simplest and stupidest thing that you can produce with this software"). Additional evaluation questions can be taken into consideration during the creative assessment process (``Doing a quick assessment of how expressive is this, how technically well done? How much does it speak to you?"). In terms of the course evaluation, the course leader can invite students' cohorts back to know more about their learning experiences (``Could ask for students or participants to kind of volunteer to stay in touch with you.") and perhaps track the students' journeys after they have completed the course (``Crawl people on the consequences of the course"). This theme reveals interesting features about how we can improve general assessments, especially those related to creative practices.
\newline
\newline
\textbf{Ethics and implications to cover when teaching AI:} When it comes to AI, ethics is a really important topic to consider, especially in a course when we are potentially introducing new people to this field. The course should cover the implications of using AI for creative practices: from IP and data set copyrights to awareness of ethics breaches in the history of creativity (``Specific creative industries IP issues, right? Because um it's certainly something which impacts a lot of people in creative industries"). The course could benefit from the support of a particular case law (``So there is an existing case law that could support an existing course.") and should cover implications of duplicated work and plagiarism (``The likelihood of it coming up or repeating simply another."). Finally, part of the responsibility of ethics should be in the interest of students in the sense that they should be sensible when collaborating with AI (``We're teaching a course like this is their responsibility to revise at least some ethics in tandem with this").

\end{document}